\documentclass[10pt,reqno,oneside]{amsart}
\usepackage{amssymb}
\usepackage[reqno]{amsmath}

\begin{document}
\numberwithin{equation}{section}

\title[Nonsymmetric Gravitational Theories in Hamiltonian Form]{A Hamiltonian 
Formulation\\ of \\Nonsymmetric Gravitational Theories}
\author{M. A. Clayton}
\address{Department of Physics, University of Toronto, Toronto, \textsc{on}, 
Canada, M5S 1A7}
\email{clayton@medb.physics.utoronto.ca}
\date{\today}
\thanks{UTPT-95-20} 
\thanks{gr-qc-9509028}

\begin{abstract}
The dynamics of a class of nonsymmetric gravitational theories is presented in 
Hamiltonian form.
The derivation begins with the first-order action, treating the generalized 
connection coefficients as the canonical coordinates and the densitised 
components of the inverse of the fundamental tensor as conjugate momenta.
The phase space of the symmetric sector is enlarged compared to the 
conventional treatments of General Relativity (GR) by a canonical pair that 
represents the metric density and its conjugate, removable by imposing 
strongly an associated pair of second class constraints and introducing Dirac 
brackets.
The lapse and shift functions remain undetermined Lagrange multipliers that 
enforce the diffeomorphism constraints in the standard form of the NGT 
Hamiltonian.
Thus the dimension of the physical constraint surface in the symmetric sector 
is not enlarged over that of GR.
In the antisymmetric sector, all six components of the fundamental tensor 
contribute conjugate pairs for the massive theory, and the absence of 
additional constraints gives six configuration space degrees of freedom per 
spacetime point in the antisymmetric sector.
For the original NGT action (or, equivalently, Einstein's Unified Field 
Theory), the $\mathrm{U}(1)$ invariance of the action is shown to  remove one 
of these antisymmetric sector conjugate pairs through an additional first 
class constraint, leaving five degrees of freedom.
The restriction of the dynamics to GR configurations is considered, as well as 
the form of the surface terms that arise from the variation of the Hamiltonian.
In the resulting Hamiltonian system for the massive theory, singular behavior 
is found in the relations that determine some of the Lagrange multipliers near 
GR and certain NGT spacetimes.
What this implies about the dynamics of the theory is not clearly understood 
at this time.
\end{abstract}

\maketitle


\section*{Introduction}

The Hamiltonian formulation of GR (viewing the metric and the extrinsic 
curvature of a spacelike hypersurface evolving in time as a first order system) 
has become popular in GR in recent years primarily due to various attempts to 
canonically quantise GR. 
However the importance of the Hamiltonian picture goes beyond this formal 
procedure.
In any classical field theory, the Hamiltonian formulation clarifies many 
dynamical issues of the system.
The Bergmann-Dirac canonical analysis begins with a classical action, tells one 
whether it is consistent, and identifies the constraints and allowed choices of 
gauge.
This identifies the physical constraint surface in phase space, the dimension of 
which gives the number of dynamical degrees of freedom in the system under 
consideration.
The resulting first-order system (Hamilton's equations) is then ideally suited 
for numerical investigations of the initial value problem.
As little is known of a general nature about solutions of the field equations in 
NGT and phenomenology tends to rely heavily on the properties of the static 
spherically symmetric solutions\footnote{\cite{Vanstone:1962} gives the general 
form of the spherically symmetric solution within the context of Einstein's 
Unified Field Theory, a special case of which was popular in the early works of 
NGT \cite{Moffat:1990}; another is the Wyman solution \cite{Wyman:1950}, 
recently resurrected for both the `old' \cite{Cornish+Moffat:1994} and massive 
\cite{Cornish:1994} versions of NGT}, it is hoped that some results of a more 
general nature will be accessible through Hamiltonian methods, or at the very 
least numerical investigations.

In the case at hand, one is looking for a description of a class of Nonsymmetric 
Gravitational Theories (NGT will refer either to this type of theory in general 
or to the `old', or massless theory, whereas mNGT will refer to the massive 
theory recently introduced in \cite{Moffat:1994, Moffat:1995b, Clayton:1995}) 
that will tell one which fields in the fundamental tensor are involved in the 
dynamical evolution of the system, and which (if any) are given by constraints 
on the initial Cauchy surface or left as freely specifiable Lagrange 
multipliers.
The analysis performed here will treat the action for NGT in first-order form, 
treating the symmetric and antisymmetric connection coefficients as canonical 
coordinates, the components of the spatial part of the fundamental tensor 
showing up as (weakly equivalent to) the conjugate momenta.
Hamilton's equations for the system are then similar to those of GR given in 
\cite{ADM:1959}, with phase space suitably enlarged to accommodate the 
antisymmetric sector, and the algebraic compatibility conditions left unimposed.
Although it is possible in principle to treat the action as a second-order 
system (depending only on the fundamental tensor and its derivatives) by 
generalizing the inversion formula of Tonnelat 
\cite{Tonnelat:1955,Tonnelat:1982}, the algebraic complications inherent in the 
solution of the generalized compatibility conditions make this approach 
unfeasible.

The action for mNGT \cite{Moffat:1994, Moffat:1995b, Clayton:1995} has recently 
been introduced in order to make the antisymmetric sector well-behaved when 
considered as a perturbation of a GR background.
The necessity of making this shift away from the original formulation of NGT was 
hinted at in some early works in NGT \cite{Mann+Moffat:1982, 
Kunstatter+Leivo+Savaria:1984}, but made a necessity by the work of Damour, 
Deser and McCarthy \cite{Damour+Deser+McCarthy:1992, Damour+Deser+McCarthy:1993} 
in showing the bad asymptotic behavior of NGT perturbations about asymptotically 
flat GR backgrounds (reviewed in \cite{Clayton:1995}).
The action for mNGT is designed so that the skew perturbation equations have the 
form of massive Kalb-Ramond theory on a GR background, with additional curvature 
coupled potential terms.
Thus in the case where the GR background is fixed, one finds three propagating 
degrees of freedom in the skew sector, and three algebraic relations that couple 
the remaining modes locally to the source (and background curvature), 
effectively removing what would be negative energy modes had they propagated.
The results found here will indicate that in general, all six antisymmetric 
components of the fundamental tensor exist as independent degrees of freedom in 
the theory.

The bulk of this paper (Sections \ref{sect:surface}-\ref{sec:Constraint 
Analysis}) will be devoted to developing the formalism and recasting the 
dynamics of NGT in Hamiltonian form.
This development is not particularly elegant, as the Hamiltonian is burdened by 
a large number of Lagrange multipliers, and the familiar patterns of the GR 
analysis become obscured.
Where possible, parallels to the analogous GR analysis have been made, in order 
to not lose sight of the (in some ways very simple) overall structure.
Indeed, although the algebra is sometimes rather messy, the analysis presented 
here is a relatively straightforward application of the Bergmann-Dirac 
constraint algorithm.

Upon concluding the constraint analysis, in Section \ref{sec:limits} the number 
of configuration space degrees of freedom per spacetime point in the old version 
of NGT is shown rigorously to be five, clearing up some confusion that has 
occurred in the literature \cite{Kunstatter+Leivo+Savaria:1984}.
This however is not a new result, as it has long been known in the context of 
Einstein's Unified Field Theory \cite{Lichnerowicz:1955}.
(Their analysis is not altered by the reinterpretation of the skew sector or the 
introduction of the source coupling into the action that defines NGT 
\cite{Moffat:1990}.)
The Cauchy analysis performed in the aforementioned works also determined that 
the Unified Field Theory (hence also NGT) `suffers' from multiple light cones in 
that there are three causal boundaries within which different modes of the 
gravitational system propagate.
It is expected that these `causal metrics' will be slightly different for 
massive NGT, and in fact the considerations of Section \ref{sect:Constraint 
problems} would seem to indicate that at least one of these has no pure GR limit 
(in the sense that as $\mathrm{g}_{[AB]}\rightarrow 0$, the related causal 
metric also vanishes).
The presence of more than one physical light cone confuses the issue of what 
constitutes the proper description of a Cauchy surface, and is discussed in more 
detail in Section \ref{Metric and Inverse}.

\section{The Action and Field Equations}
\label{sect:MNGT}

The formalism developed in Section 3 of \cite{Clayton:1995} allows one to easily 
write the action for theories with a non-symmetric `metric' and connection 
coefficients in a surface adapted coordinate system suitable for the Hamiltonian 
analysis presented here.
All covariant derivatives and curvatures have been defined with respect to a 
torsion-free connection $\Gamma^A_{BC}$ (which in general is not required to be 
compatible with any tensor in the theory), whereas what is normally treated as 
the antisymmetric part of the connection coefficients is considered as an 
additional tensor $\Lambda^A_{BC}$.
Despite the fact that this split introduces more objects into the formalism, the 
reduction to GR is made slightly more transparent, and allows the antisymmetric 
part of the torsion-free connection to be identified with the structure 
constants of a general basis in the standard way\footnote{If one had instead 
defined a covariant derivative that was not torsion-free as 
in\cite{Hlavaty:1958}, one would have to introduce additional fields into the 
action, and the split between the effects of torsion and those due to the choice 
of coordinates would be obscured.}.
Although the connection will not initially be assumed torsion-free in the 
action, a tensor of Lagrange multipliers $L^{AB}_C$ will be introduced to ensure 
vanishing torsion at the level of the field equations.
In fact this is not necessary, as one could impose the torsion-free conditions 
from the outset and remove the appropriate connection coefficients from the 
action.
This in fact is what will be done for most of the torsion-free conditions, as a 
judicious choice of arbitrary torsion terms added into the action causes the 
related connection components to disappear altogether.
However, there are three conditions that have not been imposed in this way, as 
it was felt that the formalism was simplified slightly by not doing so 
(discussed further in Section \ref{sect:Surf D A}).

The action given in \cite{Clayton:1995} ($16\pi G=c=1$) extended by the above 
mentioned Lagrange multiplier term is then\footnote{For the translation of the 
conventions used here for a coordinate basis with that of 
\cite{Moffat:1990,Moffat:1994}, see Section 1 of \cite{Clayton:1995}.}:
\begin{equation}
\label{eq:NGTAction}
S_{\textsc{ngt}}=\int_\mathbf{M} d^4\!x\,E\bigl[
-\mathbf{g}^{AB}R_{AB}^{\textsc{ns}}
-\mathbf{g}^{AB}\nabla_{e_{[A}}[W]_{_{B]}}
+\tfrac{1}{2}\alpha\mathbf{g}^{(AB)}W_A W_B
+\mathbf{l}^A\Lambda_A
+\tfrac{1}{4}m^2\mathbf{g}^{[AB]}\mathrm{g}_{[AB]}
+\mathbf{L}^{AB}_CT^C_{AB}\bigr].
\end{equation}
Each contribution to this will be denoted by corresponding Lagrangians (in order 
of appearance), each to be separately decomposed in Section \ref{sect:Surf D A}: 
${\mathcal L}_R^{\textsc{ns}}, {\mathcal L}_{\nabla W}, {\mathcal L}_{W^2}, 
{\mathcal L}_{l}, {\mathcal L}_{m}, {\mathcal L}_T$.
Densities with respect to the determinant of the fundamental tensor 
$\mathrm{g}_{AB}$, will be indicated by boldface (eg. 
$\mathbf{g}^{AB}:=\sqrt{-\text{det}[\mathrm{g}]}\mathrm{g}^{AB}$).
The action (\ref{eq:NGTAction}) encompasses that for mNGT when $\alpha=3/4$, 
`old' NGT ($\alpha=0, m=0$) (equivalent to Einstein's Unified Field Theory), and 
recovers GR in the limit that all antisymmetric components of the fundamental 
tensor are set to zero.
For simplicity, and to avoid the ambiguities inherent in defining the 
antisymmetric contributions to the matter stress-energy tensor, source-free 
spacetimes will be dealt with exclusively in this work.

The Ricci-like tensor in the action has been split up into two contributions: 
$R_{AB}^{\textsc{ns}}=R_{AB}+R^\Lambda_{AB}$.
The first is identified as the Ricci tensor (i.e. it reduces to the GR Ricci 
tensor in the limit of vanishing antisymmetric sector):
\begin{subequations}
\begin{equation}\label{eq:ncbRicci}
R_{AB}=e_C[\Gamma^C_{BA}]
-e_B[\Gamma^C_{CA}]
-\tfrac{1}{2}e_A[\Gamma^C_{BC}]
+\tfrac{1}{2}e_B[\Gamma^C_{AC}]
+\Gamma^E_{BA}\Gamma^C_{CE}
-\Gamma^E_{CB}\Gamma^C_{EA}
+\Gamma^E_{[AB]}\Gamma^C_{EC},
\end{equation}
and the second contains contributions from the antisymmetric tensor field 
$\Lambda^A_{BC}$ ($\Lambda_A:=\Lambda^B_{AB}$):
\begin{equation}\label{ncbRNgt}
R^\Lambda_{AB}=
\nabla_{e_C}[\Lambda]^C_{AB}
+\nabla_{e_{[A}}[\Lambda]_{_{B]}}
+\Lambda^C_{AD}\Lambda^D_{BC}.
\end{equation}
\end{subequations}
The Ricci tensor is defined from the two independent contractions of the 
geometric curvature tensor derived in the standard way from the connection that 
defines parallel transport of the general basis vectors \cite{Clayton:1995} by: 
$\nabla_{e_A}[e]_B=\Gamma^C_{AB}e_C$.
The basis vectors are related in the usual way to a coordinate basis via a 
vierbein $e_A={E_A}^{\mu}\partial_\mu$, 
although the vierbein in this case is not necessarily mapping the coordinate 
basis into a Lorentz frame.
As mentioned above, the torsion tensor:
\begin{equation}\label{ncbtorsion}
T^A_{BC}
=\theta^A\left[\nabla_{e_B}[e]_C-\nabla_{e_C}[e]_B-[e_B,e_C]\right]
=\Gamma^A_{BC}-\Gamma^A_{CB}-{C_{\!BC}}^A,
\end{equation}
will be constrained to vanish by variation of the Lagrange multipliers 
$L^{AB}_C$, allowing one to determine the antisymmetric part of the connection 
from the vierbeins.

Variation of both the connection coefficients ($\Gamma^A_{BC}$) and the 
antisymmetric tensor ($\Lambda^A_{BC}$) will result in the compatibility 
conditions:
\begin{equation}\label{eq:compat}
{\mathfrak C}_C^{AB}:=\nabla_{e_C}[\mathbf{g}]^{AB}
-\mathbf{g}^{DB}\Lambda^A_{CD}
-\mathbf{g}^{AD}\Lambda^B_{DC}
+\tfrac{2}{3}\alpha\delta^{[A}_C\delta^{B]}_D\mathbf{g}^{(DE)}W_E=0,
\end{equation}
where equations (4.14) and (4.28) 
%
%
%
%
%
%
%
%
%
%
%
%
of \cite{Clayton:1995} have been written in terms of the densitised components 
of the inverse of the fundamental tensor.
Note that the last term is precisely what is necessary in order for the 
compatibility conditions to be consistent with $\Lambda_A=0$.
The remaining field equations are\footnote{Note that one of (\ref{mNGTF:b}) and 
(\ref{mNGTF:c}) is redundant, as $l^A$ has been replaced in the compatibility 
conditions derived directly from the action using (\ref{mNGTF:d}), to give the 
form (\ref{eq:compat}).}:
\begin{subequations}
\label{eq:mNGT field equations}
\begin{gather}
\label{mNGTF:b}
\Lambda_A=0, \\
\label{mNGTF:d}
{\mathbf l}^A=\tfrac{1}{3}\alpha{\mathbf g}^{(AB)}W_B,\\
\label{mNGTF:c}
\nabla_{e_B}[{\mathbf g}]^{[AB]}=\alpha{\mathbf g}^{(AB)}W_B,\\
\label{mNGTF:a}
{\mathcal 
R}_{AB}=R^{\textsc{ns}}_{AB}+\nabla_{e_{[A}}[W]_{_{B]}}-\tfrac{1}{2}\alpha 
W_{A}W_{B}
-\tfrac{1}{4}m^2M_{AB}=0, 
\end{gather}
\end{subequations}
where the mass tensor appearing in the last of these is:
\begin{equation}\label{equation:mass tensor}
M _{AB}=\mathrm{g}_{[AB]}-\mathrm{g}_{CA}\mathrm{g}_{BD}\mathrm{g}^{[CD]}
+\tfrac{1}{2}\mathrm{g}_{BA}\mathrm{g}^{[CD]}\mathrm{g}_{[CD]}.
\end{equation}
Neither (\ref{eq:compat}) nor any of (\ref{eq:mNGT field equations}) will be 
used in order to pass to the second-order form of the action; they will instead 
be used in surface decomposed form to check  that Hamilton's equations (as 
derived from the NGT Hamiltonian developed in the following section) are 
equivalent to the Euler-Lagrange equations quoted here.

To date there has been no complete physical interpretation of the antisymmetric 
structure, nor will anything of the sort be attempted here.
The action (\ref{eq:NGTAction}) will be considered as the fundamental starting 
point of the Hamiltonian analysis, and the goal of the present line of research 
is be to make some headway towards understanding the dynamics of the system.
Thus in general one is left with many ambiguities as to how to make measurements 
on the skew sector, and the meaning of the various additional antisymmetric 
tensors as well as the additional contributions to the symmetric sector (GR) 
quantities is unclear.
However, once one understands what data is configurable (Cauchy data), the field 
equations fix the evolution uniquely (up to diffeomorphisms), allowing one to 
determine the physical implications of the presence of the antisymmetric sector 
in cases where the interpretation of the symmetric sector is essentially that of 
GR.
This will be the case in the asymptotic region of spacetimes that are 
asymptotically dominated by the symmetric sector, as well as in perturbative 
scenarios (discussed in Section \ref{sect:Constraint problems}) where initial 
data is close to GR configurations.

The next step in the analysis is to introduce the foliation of the spacetime 
manifold $\mathbf{M}$, and decompose all objects into components perpendicular 
and parallel to the surfaces.

\section{Metric and Compatibility in a Surface Compatible Basis}
\label{sect:surface}

Normally one would assume at this point that spacetime is globally hyperbolic, 
so that a Cauchy surface $\Sigma_0$ exists in $\mathbf{M}$ (a closed achronal 
set without edge, intersected by any smooth causal curve that is future and past 
inextensible).
Spacetime may then be viewed as the evolution of 3-geometries, that is, 
$(\mathbf{M},\mathrm{g})$ emerges from the set of field configurations on 
spacelike hypersurfaces that make up a foliation of spacetime: 
$\{(\Sigma_t,\mathrm{g}_t)|t\in I\subset \mathbb{R}^1\}$.
Then one may show that diffeomorphically equivalent initial data on $\Sigma_0$ 
generate spacetimes that are diffeomorphically equivalent, and a physical 
spacetime is then identified as an equivalence class of such solutions 
(equivalent up to diffeomorphisms).

For NGT however, a sensible definition of hyperbolicity is somewhat more 
complicated.
In Chapter IV of \cite{Maurer-Tison:1959}, it is demonstrated that the field 
equations of Einstein's Unified Field Theory propagate information along three 
separate light cones (this also occurs when considering covariant wave equations 
for particles of spin greater than or equal to one \cite{Velo+Zwanzinger:1969}).
These cones are defined by symmetric metric tensors, the set of which will be 
denoted: $\{\mathrm{g}_c\}$, and referred to as the causal metrics of 
NGT\footnote{In the Unified Field Theory, $\{\mathrm{g}_c\}=\{l,h,\gamma\}$ 
where $l$ represents $\mathrm{g}^{(\mu\nu)}$ and its inverse,
$h$ represents $\mathrm{g}_{(\mu\nu)}$ and its inverse, and $\gamma$ and its 
inverse are defined by 
$\gamma^{\alpha\beta}=\tfrac{2h}{g}h^{\alpha\beta}-l^{\alpha\beta}$ where 
$h={\rm det}[h_{\mu\nu}]$ and $\mathrm{g}={\rm det}[\mathrm{g}_{\mu\nu}]$.
These metrics were found to be compatible in the sense that there is a largest 
and smallest speed of light, and all three metrics merge into the Riemannian 
metric of GR in the limit of vanishing antisymmetric sector.
As the set of causal metrics is not known explicitly for mNGT at this time, 
strictly speaking one cannot discuss the causal structure of any given 
spacetime.}.
Thus the statement that a vector is timelike (spacelike) with respect to 
$\{\mathrm{g}_c\}$ will indicate that it is timelike (spacelike) with respect to 
all of the causal metrics in $\{\mathrm{g}_c\}$.
Global hyperbolicity would then require that $\Sigma_0$ be achronal with respect 
to $\{\mathrm{g}_c\}$, and the causal curves would of course also be causal with 
respect to $\{\mathrm{g}_c\}$.
This would ensure that none of $\{\mathrm{g}_c\}$ would become degenerate 
anywhere in spacetime, and hence that $\Sigma_0$ is in fact a Cauchy surface.
At this stage it is not known whether one can require such a condition on the 
fundamental tensor\footnote{What is commonly referred to as the `metric' of NGT 
is the tensor determined directly by the field equations (appearing in 
(\ref{eq:NGTAction})), the determinant of which defines the spacetime volume 
element in the action, and from which one derives the causal metrics 
$\{\mathrm{g}_c\}$.
The presence of antisymmetric components indicates that it is in fact not a 
metric (in the usual sense) at all, and will therefore be referred to as the 
`fundamental tensor' (in accordance with Lichnerowicz \cite{Lichnerowicz:1955}).
Note however that one may consider it as a Hermitian metric in a complex or 
hyperbolic complex space 
\cite{Crumeyrolle:1967,Kunstatter+Yates:1981,Kunstatter+Moffat+Malzan:1983}, or 
in more general spaces \cite{Mann:1984}.}.
However even if this is not possible, the Hamiltonian system is still relevant 
for considering the dynamics of the theory locally.
One may instead specify initial data on a large enough $S_0\subset\Sigma_0$ in 
order that the configuration on $S_t\subset\Sigma_t$ ($S_t$ is a closed achronal 
set) is determined uniquely.
This requires that $\{\mathrm{g}_c\}$ is known in order to determine $D^+(S_0)$ 
or $D^-(S_t)$ (the future (past) Cauchy development or domain of dependance of 
$S_0$ ($S_t$)), to ensure that $S_0\subset D^+(S_0)$ or $S_0\supset D^-(S_t)$.
For the purposes of this work, it will suffice to discuss the global problem, as 
one may restrict the results to in a straightforward manner once the causal 
metrics of mNGT are known.

\subsection{The Surface Adapted Basis}
\label{Metric and Inverse}

We begin by assuming that there is a time function $t$ that is used to foliate 
$\mathbf{M}$ into hypersurfaces of constant time $\Sigma_t$.
This requires that each (3-)surface of constant time, $\Sigma_t$, be spacelike 
with respect to $\{\mathrm{g}_c\}$, and $\nabla_\mathsf{t}t=1$ defines a vector 
$\mathsf{t}$ that is timelike with respect to $\{\mathrm{g}_c\}$.
This ensures that the degrees of freedom that travel along each of the light 
cones in $\{\mathrm{g}_c\}$ will evolve forward in what has been chosen as time.

In GR, one may then introduce a coordinate basis on the surface and a unit 
normal, such that the metric is of the form: 
$\mathrm{g}=\theta^\perp\otimes\theta^\perp-\gamma_{ab}\theta^a\otimes\theta^b$.
This effectively reduces the metric to a Riemannian metric on $\Sigma$, and the 
projection of $dt$ perpendicular and parallel to the surface are the lapse and 
shift functions respectively \cite{Isenberg+Nester:1980}.
It is the components of this metric on $\Sigma$ that are taken to be the 
Canonical coordinates in the Hamiltonian approach to GR, and all other spacetime 
tensors are decomposed onto the surface by considering the components 
perpendicular and parallel to the surface as separate tensors on $\Sigma$.

On attempting to generalize this to NGT, one finds that the presence of more 
than one physical metric on $\mathbf{M}$ implies that there is no physically 
well-motivated or natural choice of metric on $\Sigma$ that would play the same 
role.
One may introduce a coordinate basis on the surface in the usual way, but in 
general each of the metrics in $\{\mathrm{g}_c\}$ will provide a different 
definition of the unit normal vector $e_\perp$, and the resulting decomposition 
of tensors onto $\Sigma$ will be inequivalent.
However as this is merely a choice of  parameterization, the physical content of 
the system will be independent of the choice of metric used to define the 
surface decomposition.
In this work, the components of $\mathrm{g}^{(AB)}$ will be used to define the 
unit normal $e_\perp$, therefore making a particular choice of decomposition:
\begin{subequations}\label{eq:basis}
\begin{gather}
\label{eq:basis:b}
\mathrm{g}^{-1}=e_\perp\otimes e_\perp
+B^ae_\perp\otimes e_a-B^ae_a\otimes e_\perp
-\gamma^{ab}e_a\otimes e_b,\\
\label{eq:basis:a}
\mathrm{g}^{-1}_{\text{symm}}=e_\perp\otimes e_\perp-
\gamma^{(ab)}e_a\otimes e_b,
\end{gather}
\end{subequations}
where `symm' indicates the symmetric part of the inverse of the fundamental 
tensor. 
This choice is made in order to simplify as much as possible the form of the 
action (\ref{eq:NGTAction}), as the fundamental tensor appears in the form 
$\mathrm{g}^{AB}$.

The time vector is decomposed as $\mathsf{t}=t^Ae_A=Ne_\perp
+N^ae_a$, where $(N,N^a)$ are the lapse (component of $\mathsf{t}$
along $e_\perp$) and shift (projection of $\mathsf{t}$ on the
surface $\Sigma_t$).
This basis can therefore be written in terms of the coordinate
basis ($\partial_t,\partial_a$) by:
$e_\perp=\frac{1}{N}\partial_t-\frac{N^a}{N}\partial_a$, $e_a=\partial_a$, and 
the covector bases by:
$\theta^\perp=Ndt$, $\theta^a=dx^a+N^adt$.
This defines the vierbein through $\theta^A={E^A}_\mu dx^\mu$, the determinant 
of which is:
\begin{equation}
E:=\det[{E^A}_\mu ]=N.
\end{equation}
Note that $e_a$ is a coordinate basis on $\Sigma$ and can be therefore be 
thought of as acting on tensor fields as an ordinary derivative $\partial_a$. 
It is also useful to denote surface vectors (not just the components) by 
$\vec{V}:=V^ae_a$ and covectors by $\vec{\omega}:=\omega_a\theta^a$.
It is not difficult to check that in the coordinate basis 
$(\partial_t,\partial_a)$, $\mathrm{g}^{-1}_{\text{symm}}$ and its inverse take 
on the usual ADM form \cite{ADM:1959}.

The fundamental tensor is found from the inverse of (\ref{eq:basis:b}):
\begin{subequations}\label{eq:fmet}
\begin{equation}
\mathrm{g}=F\theta^\perp\otimes\theta^\perp
-(\alpha_a-\beta_a)\theta^\perp\otimes\theta^a
-(\alpha_a+\beta_a)\theta^a\otimes\theta^\perp
-G_{ab}\theta^a\otimes\theta^b,
\end{equation}
where:
\begin{equation}\label{invmetcom}
F:=1+G_{(ab)}B^aB^b,\quad 
\alpha_a:=G_{[ab]}B^b,\quad
\beta_a:=G_{(ab)}B^b,
\end{equation}
from which one finds $\alpha_aB^a=0$.
The spatial part of the fundamental tensor ($G_{ab}$) is given as the inverse of 
$\gamma-B\otimes B$:
\begin{equation}\label{eq:ginvG}
(\gamma^{ac}-B^a B^c)G_{cb}=
G_{bc}(\gamma^{ca}-B^c B^a)=
\delta^a_b.
\end{equation}
\end{subequations}
Note that invertibility of the fundamental tensor (\ref{eq:fmet}) is required in 
order that the volume element in the action be nondegenerate, and this requires 
that $G_{ab}$ exist.
Given that the fundamental tensor is a nondegenerate solution of the field 
equations in no way guarantees that $\{\mathrm{g}_c\}$ are all nondegenerate, 
allowing for the possibility that a regular solution of the field equations has 
regions in spacetime where NGT perturbations `see' a singular spacetime.

Some useful identities may be derived from (\ref{eq:ginvG}):
\begin{equation}
\begin{split}
\gamma^{(ac)}G_{[cb]}+ \gamma^{[ac]}G_{(cb)}+B^a\alpha_b=0,&\quad
\gamma^{(ac)}G_{(cb)}+ \gamma^{[ac]}G_{[cb]}-B^a\beta_b=\delta^a_b,\\
\gamma^{(ac)}\alpha_c+\gamma^{[ac]}\beta_c=0,&\quad
\gamma^{(ac)}\beta_c+\gamma^{[ac]}\alpha_c=FB^a.
\end{split}
\end{equation}
Varying $\sqrt{-\mathrm{g}}$ with respect to the densitised components of the 
inverse of the fundamental tensor results in:
\begin{equation}\label{eqn:variation density}
\delta\sqrt{-\mathrm{g}}=-\frac{2}{2-F}\beta_a\delta\mathbf{B}^a
+\frac{1}{2-F}G_{ba}\delta\pmb{\gamma}^{ab}.
\end{equation}
This will be used order to compute the variation of the Hamiltonian of NGT with 
respect to the same densitised components, as well as applied as an ordinary 
derivative relation in the surface-decomposed compatibility conditions in 
Appendix \ref{sec:compat}.
Note that the condition $F=2$ corresponds to the case where $\sqrt{-\mathrm{g}}$ 
is independent of $(\mathbf{B}^a, \pmb{\gamma}^{ab})$, and $F\neq 2$ will be 
assumed throughout this work. 

The non-vanishing structure constants for this basis are found to be:
\begin{equation}\label{Cs}
\begin{split}
[e_\perp,e_a]&={C_{\!\perp a}}^\perp e_\perp+{C_{\!\perp a}}^b e_b,  \\
{C_{\!\perp a}}^\perp&=e_a[\ln(N)],\quad
{C_{\!\perp a}}^b=\frac{1}{N}e_a[N^b],
\end{split}
\end{equation}
and if one were to require that the connection $\Gamma$ be torsion-free using 
(\ref{ncbtorsion}), the antisymmetric components of the connection would be:
\begin{equation}\label{Gamskew}
\Gamma^\perp_{[\perp a]}=\frac{1}{2}e_a[\ln(N)],\quad
\Gamma^b_{[\perp a]}=\frac{1}{2N}e_a[N^b], \quad 
\Gamma^\perp_{[ab]}=0,\quad
\Gamma^a_{[bc]}=0.
\end{equation}
A final useful property of this basis is:
\begin{equation}\label{eq:Jacobi}
e_b[{C_{\!a\perp}}^b]+e_a[{C_{\!\perp b}}^b]+
{C_{\!a\perp}}^\perp{C_{\!b\perp}}^b+{C_{\!a\perp}}^b{C_{\!\perp b}}^\perp=0,
\end{equation}
which can either be derived directly from a contraction of the Jacobi identity, 
or proved by a brute force insertion of the structure constants (\ref{Cs}).
This will be useful when identifying the diffeomorphism constraints, linking 
them to the algebraic field equations in Appendix \ref{sec:feq}.  

In beginning with the components of the inverse of the fundamental tensor in 
(\ref{eq:basis}), the configuration of the system is being described by the 
degrees of freedom $N, N^a,\mathbf{B}^a,\pmb{\gamma}^{ab}$ (the latter two are 
included as densities since they will turn up later on as weakly equal to the 
conjugate momenta in this form).
As one can see from the form of (\ref{eq:fmet}), different parameterizations of 
the fundamental tensor will have a large effect on the details of the analysis 
presented here.
If one had begun by requiring that the symmetric part of the fundamental tensor 
took on the ADM form, then the inverse (that appears in the action) would have 
had nontrivial $\mathrm{g}^{(0a)}$ components, and the analysis would have been 
more complicated.
In NGT, one must be slightly more careful to be consistent in what one is 
calling the independent degrees of freedom, as the antisymmetric degrees of 
freedom will mix with the symmetric sector as one moves from spatial components 
of the fundamental tensor to the spatial components of its inverse.

\subsection{Surface Decomposition}

Four-dimensional covariant derivatives are decomposed as usual 
\cite{Isenberg+Nester:1980} into surface covariant derivatives (written as 
$\nabla^{(3)}_a$) whose action on basis vectors in $T\Sigma$ is given by 
$\nabla^{(3)}_a[e]_b=\Gamma^c_{ab}e_c$, and contributions from derivatives off 
of $\Sigma$ which will be written in terms of the surface tensors:
\begin{equation}
\begin{split}
\Gamma:&=\Gamma^\perp_{\perp\perp},  \\
c_a:=\Gamma^\perp_{a\perp},\quad
a_a:&=\Gamma^\perp_{\perp a},\quad
\sigma^a:=\Gamma^a_{\perp\perp},  \\
w^a_{\;b}:=\Gamma^a_{\perp b},\quad
u^a_{\;b}:&=\Gamma^a_{b\perp}, \quad 
k_{ab}:=\Gamma^\perp_{ab}.
\end{split}
\end{equation}
(Both $k$ and $\Gamma^c_{ab}$ will initially be considered as nonsymmetric, 
allowing for the possibility of non-zero torsion.
The torsion-free conditions will be imposed for these components very early in 
the development, and from then on they will be assumed to be symmetric.)
In GR most of these are related by algebraic compatibility conditions 
($\sigma^a=\gamma^{ab}a_b, c_a=0,\Gamma=0,u^a_{\;b}=\gamma^{ac}k_{bc}$), and the 
absence of torsion would allow one to further relate $a$ and $(u,w)$ to the 
structure constants.
The skew tensor $\Lambda$:
\begin{equation}
b_a:=\Lambda^\perp_{a\perp},\quad
j_{ab}:=\Lambda^\perp_{ab},\quad
v^a_{\;b}:=\Lambda^a_{\perp b},\quad
\lambda^a_{bc}:=\Lambda^a_{bc},
\end{equation}
the torsion Lagrange multipliers:
\begin{equation}
L^a:=L^{a\perp}_\perp,\quad
L^{ab}:=L^{ab}_\perp,\quad
L^a_{\;b}:=L^{a\perp}_b,\quad
L^{bc}_a:=L^{bc}_a,
\end{equation}
and the vector fields: $W_A=(W,W_a)$ and $l^A=(l,l^a)$, are all decomposed in a 
similar fashion.
Traces are denoted as $v:=v^a_{\;a}$, and the traceless part: 
$v^a_{Tb}:=v^a_{\;b}-1/3\delta^a_b v$.
The components of $\Lambda_A$ are given by $\Lambda_\perp=v, 
\Lambda_a=b_a+\lambda^b_{ab}$, and $j_{ab}$ is an antisymmetric surface tensor 
by definition.
It also will be useful to define a few combined quantities that will appear:
\begin{subequations}
\begin{align}
K_{ab}&:=k_{ab}+j_{ab},\\
k^a_{\;b}&:=\tfrac{1}{2}(\gamma^{ac}K_{bc}+\gamma^{ca}K_{cb})
=\gamma^{(ac)}k_{bc}+\gamma^{[ac]}j_{bc},\\
j^a_{\;b}&:=\tfrac{1}{2}(\gamma^{ac}K_{bc}-\gamma^{ca}K_{cb})
=\gamma^{[ac]}k_{bc}+\gamma^{(ac)}j_{bc},\\
u^{ab}&:=\tfrac{1}{2}(\gamma^{(ac)}u^{b}_{\;c}+\gamma^{(bc)}u^{a}_{\;c}
+\gamma^{[ac]}v^{b}_{\;c}+\gamma^{[bc]}v^{a}_{\;c}),\\
v^{ab}&:=\tfrac{1}{2}(\gamma^{[ac]}u^{b}_{\;c}
-\gamma^{[bc]}u^{a}_{\;c}
+\gamma^{(ac)}v^{b}_{\;c}-\gamma^{(bc)}v^{a}_{\;c}).
\end{align}
\end{subequations}
Aside from $K_{ab}$, which will be useful as a notational tool to combine the 
results of both sectors into one relation, these combinations correspond to 
symmetric ($k_{ab},u^a_{\;b}$) and antisymmetric ($j_{ab},v^a_{\;b}$) sector 
contributions once an index has been `raised' by the spatial fundamental tensor 
$\gamma^{ab}$.

\subsection{Lie Derivatives}

The standard Lie derivatives defined on $\Sigma$:
\begin{subequations}
\begin{gather}
\pounds^{(3)}_{\vec{X}}[\vec{Y}]=[\vec{X},\vec{Y}]=(X^ae_a[Y^b]-Y^ae_a[X^b])e_b, 
\quad
\pounds^{(3)}_{\vec{X}}[Y^\perp]=X^ae_a[Y^\perp],\\
\pounds^{(3)}_{\vec{X}}[\vec{\omega}]=(X^ae_a[\omega_c]
+\omega_ae_c[X^a])\theta^c,\quad
\pounds^{(3)}_{\vec{X}}[\omega_\perp]=X^ae_a[\omega_\perp],
\end{gather}
\end{subequations}
(and the standard generalization to higher-order tensors) will be used 
throughout.
Note that the perpendicular components are treated as scalar fields as far as 
surface defined derivatives are concerned. 
This is also true of the derivative off of $\Sigma$ \cite{Isenberg+Nester:1980}:
\begin{subequations}
\begin{gather}
\mathrm{d}_{X^\perp e_\perp}[\vec{Y}]=X^\perp(e_\perp[Y^b]
+Y^a{C_{\!\perp a}}^b)e_b,\quad
\mathrm{d}_{X^\perp e_\perp}[Y^\perp]=X^\perp e_\perp[Y^\perp],\\
\mathrm{d}_{X^\perp e_\perp}[\vec{\omega}]=X^\perp(e_\perp[\omega_b]
+\omega_aC_{\!b\perp}^{\;\;\;a})\theta^b,\quad
\mathrm{d}_{X^\perp e_\perp}[\omega_\perp]=X^\perp e_\perp[\omega_\perp],
\end{gather}
\end{subequations}
where the structure constant takes into account the possibility that the surface 
basis may `move' on $\Sigma$ as one moves perpendicularly off the surface.
These definitions are also extended to arbitrary tensors in the standard manner.

From the spacetime Lie derivatives of tensor densities:
\begin{subequations}
\begin{align}
\pounds_{X}[\mathbf{T}]&=\sqrt{-\mathrm{g}}\pounds_{X}[T]
+T\pounds_{X}[\sqrt{-\mathrm{g}}],\\
\pounds_{X}[\sqrt{-\mathrm{g}}]&
=\tfrac{1}{2}\mathbf{g}^{CB}\pounds_{X}[\mathrm{g}_{BC}]
=-\tfrac{1}{2}\mathbf{g}_{CB}\pounds_{X}[\mathrm{g}^{BC}],
\end{align}
and using the defined surface Lie derivatives, one finds:
\begin{equation}
\partial_t[\sqrt{-\mathrm{g}}]=\mathrm{d}_\mathsf{t}[\sqrt{-\mathrm{g}}]
:=N\mathrm{d}_{e_\perp}[\sqrt{-\mathrm{g}}]
+\pounds^{(3)}_{\vec{N}}[\sqrt{-\mathrm{g}}]
=N(e_\perp[\sqrt{-\mathrm{g}}]
-\sqrt{-\mathrm{g}}{C_{\!\perp a}}^a)+\nabla^{(3)}_{a}[\mathbf{N}]^a.
\end{equation}
\end{subequations}
Time derivatives of arbitrary tensors or tensor densities are given as in GR by: 
$\partial_t[Z]=\mathrm{d}_{\mathsf{t}}[Z]=N\mathrm{d}_{e_\perp}[Z]+\pounds^{(3)}
_{\vec{N}}[Z]$, where $Z$ is any tensor density.

One would now like to identify the extrinsic curvature of the surface $\Sigma$ 
under consideration.
Although there are a variety of equivalent ways of describing it in the context 
of Riemannian geometry, none of these are equivalent for the type of 
nonsymmetric theory considered here.
One way is to compute the perpendicular component of the parallel transport of a 
surface vector along the surface (Gauss' formula \cite{Abraham+Marsden:1987}): 
$\theta^\perp\bigl[\nabla_{\vec{M}}[\vec{N}]\bigr]=\Gamma^\perp_{ab}M^aN^b
=k_{ab}M^aN^b$, in conjunction with the fact that the surface projection is just 
the surface transport result in GR 
$\theta^a\bigl[\nabla_{\vec{M}}[\vec{N}]\bigr]=\theta^a\bigr
[\nabla^{(3)}_{\vec{M}}[\vec{N}]\bigr]$ \cite{Choquet-Bruhat+:1989}.
One could equivalently describe it by the change in the unit normal as it is 
parallel transported along the surface (Weingarten equation) $\nabla_a[e_\perp]$ 
\cite{MTW:1973}.
In GR, the magnitude of this vector does not change when transported along the 
surface (since $\Gamma^\perp_{a\perp}=c_a=0$), and the result is a vector 
tangent to the surface describing the amount that the normal vector must 
`rotate' in order to remain normal 
$\nabla_a[e_\perp]=\Gamma^b_{a\perp}e_b=u^b_{\;a}e_b$.
Metric compatibility relates these two definitions since 
$k_{ab}=\gamma_{bc}u^c_{\;a}$.
In the case of NGT, this definition is clouded by the fact that in general 
$c_a\neq 0$, and the length of the unit normal will not be preserved under (this 
definition of) parallel transport.
One could also identify the extrinsic curvature of $\Sigma$ with the change of 
the 3-metric perpendicularly off the surface 
$\nabla_{e_\perp}[\mathrm{g}][e_a\otimes 
e_b]=-\mathrm{d}_{e_\perp}[\gamma]_{ab}+\gamma_{cb}u^c_{\;a}
+\gamma_{ac}u^c_{\;b}=0$, giving 
$\mathrm{d}_{e_\perp}[\gamma]_{ab}=2k_{ab}$ (or the equivalent derived from the 
inverse metric).

Although equivalent in GR, none of these definitions coincide in nonsymmetric 
theories in general.
Indeed, since the compatibility conditions cannot entirely be written in terms 
of a metric compatible connection, there is no truly natural choice of parallel 
transport, and any choice of `extrinsic curvature' based upon a covariant 
derivative is conventional at best.
The approach followed here will be the first of those described above, 
essentially due to convenience.
It turns out that with the choice of decomposition made here, 
$\Gamma^\perp_{ab}=k_{ab}$ is a canonical coordinate, and as such plays a more 
central role than $u^a_{\;b}$ (which will turn out to be a derived quantity, as 
it acts as a determined Lagrange multiplier), or the relation derived from the 
evolution of the spatial metric.

The stage is now set, as one now has the tools necessary to decompose spacetime 
tensors and covariant derivatives into spatially covariant objects, and identify 
time derivatives in the action.
The configuration space variables have also been chosen to be: ($N, 
N^a,\mathbf{B}^a,\pmb{\gamma}^{ab}$).
In Appendix \ref{sec:compat} the compatibility conditions (\ref{eq:compat}) are 
decomposed, and the field equations (\ref{eq:mNGT field equations}) in Appendix 
\ref{sec:feq}.
In both cases, algebraic conditions that will determine Lagrange multipliers on 
$\Sigma$ have been identified, as well as those responsible for time evolution.
The reader is again reminded that all of these will be derived from the 
Hamiltonian, and are only given here in order to introduce some defined 
quantities ($\mathfrak{C}, \mathfrak{G}, \mathfrak{Z}$) that will help to 
simplify the presentation of the algebra, and ensure that one is finding results 
that are in accord with the Lagrangian variational principle. 

\section{Determining the Hamiltonian}

This section is concerned with obtaining the form of the Hamiltonian for NGT.
The various terms in the Lagrangian (\ref{eq:NGTAction}) will be decomposed into 
surface compatible form, and the fields that appear as time derivatives 
identified.
After this, writing down the form of the Hamiltonian turns out to be a rather 
simple task, however the work is far from done.
The Hamiltonian is riddled with Lagrange multipliers, and the constraint 
analysis is where one begins to see the full structure of NGT emerge.

\subsection{The Surface Decomposed Action}
\label{sect:Surf D A}

The simplest Lagrangian density terms in (\ref{eq:NGTAction}) are decomposed as: 
\begin{subequations}
\begin{align}
\label{eq:Hm}
{\mathcal L}_m&=-\mathcal{H}_m=\tfrac{1}{2}m^2N\mathbf{B}^a\beta_a
+\tfrac{1}{4}m^2N\pmb{\gamma}^{[ab]}G_{[ab]},\\
{\mathcal L}_l&=N\mathbf{l}v^a_{\;a}+N\mathbf{l}^a(b_a+\lambda^b_{ab}),\\
{\mathcal L}_T&=2N[{\mathbf L}^{ab}K_{[ab]}+{\mathbf L}^{ab}_c\Gamma^c_{[ab]}
+{\mathbf  L}^a_{\;b}(u^b_{\;a}-w^b_{\;a}+{C_{\!\perp a}}^b)
+{\mathbf L}^a(c_a-a_a+{C_{\!\perp a}}^\perp)],\\
{\mathcal L}_{W^2}&=\tfrac{1}{2}\alpha N
\bigl[\sqrt{-\mathrm{g}}(W)^2-\pmb{\gamma}^{(ab)}W_aW_b\bigr],\\
{\mathcal L}_{\nabla W}&=N\mathbf{B}^a
\bigl[\nabla^{(3)}_{a}[W] -\mathrm{d}_{e_\perp}[W]_a+W(a_a-c_a)\bigr]
+N\pmb{\gamma}^{[ab]}\nabla^{(3)}_{{[a}}[W]_{b]}.
\end{align}
The first of these defines $\mathcal{H}_m$, which will be useful in Section 
\ref{sect:phi}.
The remaining contribution ${\mathcal L}_{R^{\textsc{ns}}}$, will be further 
split into contributions from each of the symmetric and antisymmetric sectors as 
${\mathcal L}_{R}$ and ${\mathcal L}_{R^\Lambda}$, and further split to give 
contributions from each component of the tensor (for example, ${\mathcal 
L}_{R_{(ab)}}$ corresponds to the contribution from $N\gamma^{(ab)}R_{(ab)}$).
Note that the fact that $\mathrm{g}^{(\perp a)}=0$ in this decomposition implies 
that there is no contribution from ${\mathcal L}_{R_{(\perp a)}}$, which is 
essentially why the choice of surface adapted basis (\ref{eq:basis}) was made.
Making a choice of parameterization in which $\mathrm{g}^{(\perp a)}\neq 0$ 
would have explicitly introduced a term involving $\partial_t{B}^a$ into the 
action (see (\ref{eq:Raperp})).
The remaining contributions are:
\begin{equation}
\begin{split}
{\mathcal L}_{R^\Lambda_{\perp\perp}}&=-N\sqrt{-\mathrm{g}}v^a_{\;b}v^b_{\;a},\\
{\mathcal L}_{R^\Lambda_{(ab)}}&=N\pmb{\gamma}^{(ab)}
\bigl[b_ab_b-j_{ac}v^c_{\;b}-j_{bc}v^c_{\;a}
+\lambda^d_{ac}\lambda^c_{bd}\bigr],\\
{\mathcal L}_{R^\Lambda_{[\perp a]}}&=-N\mathbf{B}^a\bigl[
\mathrm{d}_{e_\perp}[\lambda]^b_{ab}
-\mathrm{d}_{e_\perp}[b]_a
+2\nabla^{(3)}_{b}[v]^b_{\;a}
-\nabla^{(3)}_{a}[v] \\
&\quad\quad +2b_bu^b_{\;a}-2b_au
+2(a_b-c_b)v^b_{\;a}
-(a_a-c_a)v
-2u^b_{\;c}\lambda^c_{ba}
+2j_{ab}\sigma^b\bigr],\\
{\mathcal L}_{R^\Lambda_{[ab]}}&=N\pmb{\gamma}^{[ab]}\bigl[
\mathrm{d}_{e_\perp}[J]_{ab}
+\nabla^{(3)}_{c}[\lambda]^c_{ab}
+\nabla^{(3)}_{{[a}}[b]_{b]}
+\nabla^{(3)}_{{[a}}[\lambda]^c_{b]c}  \\
&\quad\quad -j_{ac}u^c_{\;b}-j_{cb}u^c_{\;a}
+(\Gamma+u)j_{ab}
-k_{ca}v^c_{\;b}+k_{cb}v^c_{\;a}
+a_ab_b-a_bb_a+a_c\lambda^c_{ab}\bigr],\\
{\mathcal L}_{R_{\perp\perp}}&=-N\sqrt{-\mathrm{g}}\bigl[
-\mathrm{d}_{e_\perp}[u]
+\nabla^{(3)}_{a}[\sigma]^a
+\Gamma u
+\sigma^a(a_a-2c_a)
-u^a_{\;b}u^b_{\;a}\bigr],\\
{\mathcal L}_{R_{(ab)}}&=N\pmb{\gamma}^{(ab)}\bigl[
R^{(3)}_{(ab)}
+\mathrm{d}_{e_\perp}[k]_{(ab)}
-\nabla^{(3)}_{{(a}}[a]_{b)}
-a_aa_b
+(\Gamma+u)k_{ab}
-k_{cb}u^c_{\;a}
-k_{ac}u^c_{\;b}\bigr],
\end{split}
\end{equation}
\end{subequations}
where $R^{(3)}_{(ab)}$ is the surface Ricci tensor given by (\ref{eq:ncbRicci}) 
determined from surface covariant derivatives, and the identity $R_{[AB]}=0$ 
(see Section IV in \cite{Clayton:1995}) has been used.
%
%
%
%
%
%
%
 
The terms that result in time derivatives in the action are then:
\begin{equation*}
-N\mathbf{B}^a\mathrm{d}_{e_\perp}[W]_a
-N\mathbf{B}^a\mathrm{d}_{e_\perp}[\lambda]^b_{ab}
+N\mathbf{B}^a\mathrm{d}_{e_\perp}[b]_a
+N\sqrt{-\mathrm{g}}\mathrm{d}_{e_\perp}[u]
+N\pmb{\gamma}^{ab} \mathrm{d}_{e_\perp}[K]_{ab},
\end{equation*}
and so defining the vector field:
\begin{equation}\label{eq:Wbar}
\overline{W}_a=-W_a-\lambda^b_{ab}+b_a,
\end{equation}
will simplify the remaining analysis.
Rewriting the above in terms of the time derivative gives:
\begin{equation*}
\begin{split}
&\mathbf{B}^a\mathrm{d}_\mathsf{t}[\overline{W}]_a
+\sqrt{-\mathrm{g}}\mathrm{d}_\mathsf{t}[u]
+\pmb{\gamma}^{ab} \mathrm{d}_\mathsf{t}[K]_{ab}\\
-&\mathbf{B}^a\pounds^{(3)}_{\vec{N}}[\overline{W}]_a
-\sqrt{-\mathrm{g}}\pounds^{(3)}_{\vec{N}}[u]
-\pmb{\gamma}^{ab} \pounds^{(3)}_{\vec{N}}[K]_{ab},
\end{split}
\end{equation*}
where:
\begin{subequations}
\begin{align}
\pounds^{(3)}_{\vec{N}}[u]&= N^a\nabla^{(3)}_{a}[u],\\
\pounds^{(3)}_{\vec{N}}[\overline{W}]_a&= 
N^b\nabla^{(3)}_{b}[\overline{W}]_a+\overline{W}_b\nabla^{(3)}_{a}[N]^b,\\
\pounds^{(3)}_{\vec{N}}[K]_{ab}&=N^c\nabla^{(3)}_{c}[K]_{ab}
+K_{cb}\nabla^{(3)}_{a}[N]^c
+K_{ac}\nabla^{(3)}_{b}[N]^c.
\end{align}
\end{subequations}
Although these have been written in surface covariant form, all dependence on 
the spatial connection coefficients vanish in these.
As it is clear that $\overline{W}_a$ will be a canonical coordinate, 
(\ref{eq:Wbar}) will henceforth be used to replace $W_a$ everywhere in the 
remainder of this work.

Gathering together terms in the action that are multiplied by the same component 
of the fundamental tensor and identifying the time derivatives as above, the 
Lagrangian density for NGT can now be written:
\begin{equation}\label{eq:L}
\begin{split}
{\mathcal L}_{\textsc{ngt}}=&
-\mathcal{H}_m
+\mathbf{B}^a\mathrm{d}_\mathsf{t}[\overline{W}]_a
+\sqrt{-\mathrm{g}}\mathrm{d}_\mathsf{t}[u]
+\pmb{\gamma}^{ab} \mathrm{d}_\mathsf{t}[K]_{ab}
-\mathbf{B}^a\pounds^{(3)}_{\vec{N}}[\overline{W}]_a
-\sqrt{-\mathrm{g}}\pounds^{(3)}_{\vec{N}}[u]
-\pmb{\gamma}^{ab} \pounds^{(3)}_{\vec{N}}[K]_{ab}\\
&+N\sqrt{-\mathrm{g}}\bigl[
lv+l^a(b_a+\lambda^b_{ab})
+2L^a(c_a-a_a+{C_{\!\perp a}}^\perp)\\
&\quad\quad +\tfrac{1}{2}\alpha(W)^2
-v^a_{\;b}v^b_{\;a}
-\nabla^{(3)}_{a}[\sigma]^a
-\Gamma u
-\sigma^a(a_a-2c_a)
+u^a_{\;b}u^b_{\;a}\bigr]\\
&+N\mathbf{B}^a\bigl[
\nabla^{(3)}_{a}[W]+W(a_a-c_a)
-2\nabla^{(3)}_{b}[v]^b_{\;a}
+\nabla^{(3)}_{a}[v] \\
&\quad\quad -2b_bu^b_{\;a}
+2b_au
-2(a_b-c_b)v^b_{\;a}+(a_a-c_a)v
+2u^b_{\;c}\lambda^c_{ba}
-2j_{ab}\sigma^b\bigr]\\
&+N\pmb{\gamma}^{[ab]}\bigl[
-\nabla^{(3)}_{{[a}}[\overline{W}]_{b]}
+\nabla^{(3)}_{c}[\lambda]^c_{ab}
+2\nabla^{(3)}_{{[a}}[b]_{b]} \\
&\quad\quad -j_{ac}u^c_{\;b}-j_{cb}u^c_{\;a}
+(\Gamma+u)j_{ab}
-k_{ca}v^c_{\;b}+k_{cb}v^c_{\;a}
+a_ab_b-a_bb_a+a_c\lambda^c_{ab}\bigr]\\
&+N\pmb{\gamma}^{(ab)}\bigl[
R^{(3)}_{(ab)}
-\nabla^{(3)}_{{(a}}[a]_{b)}
-a_aa_b  
+(\Gamma+u)k_{ab}
-k_{cb}u^c_{\;a}
-k_{ac}u^c_{\;b}\\
&\quad\quad-\tfrac{1}{2}\alpha(\overline{W}_a+\lambda^c_{\;ac}-b_a)
(\overline{W}_b+\lambda^d_{\;bd}-b_b)
+b_ab_b-j_{ac}v^c_{\;b}-j_{bc}v^c_{\;a}
+\lambda^d_{ac}\lambda^c_{bd}\bigr].
\end{split}
\end{equation}
Up to this point, the torsion-free conditions have been applied arbitrarily in 
order to simplify the Lagrangian density as much as possible.
At this stage, the fields $k_{[ab]}$ and $\Gamma^a_{[bc]}$ have been dropped 
along with their associated torsion Lagrange multipliers $L^{ab}$ and 
$L^{ab}_c$, as they would appear only in these torsion Lagrange multiplier 
terms.
From this point onwards, both $k_{ab}$ and $\Gamma^c_{ab}$ will refer to 
symmetric objects.
The Lagrange multiplier term imposing the torsion-free condition:
\begin{equation}\label{eq:wcondition}
w^a_{\;b}= u^a_{\;b}+{C_{\!\perp b}}^a,
\end{equation}
has also been dropped, as once all terms (in the action as well as compatibility 
conditions and field equations) are written in terms of $\mathrm{d}_{e_\perp}$, 
$w^a_{\;b}$ will not appear either, and may be determined by 
(\ref{eq:wcondition}) once $u^a_{\;b}$ is known.
One would also like to impose the torsion conditions to solve for $a_a$ 
explicitly, however this would introduce a number of terms involving the 
structure constants ${C_{\!\perp a}}^\perp$ into the action, compatibility 
conditions and field equations, and it was felt that avoiding this simplified 
the algebra.
Although there is nothing preventing one from doing this, the presentation is 
slightly simpler if the torsion condition is not imposed in this case.

\subsection{Legendre Transform}
\label{subsect:ham}

In the form (\ref{eq:L}), the fields with conjugate momenta are trivial to 
identify (these are densities by definition, and will not be written in 
boldface)\footnote{Note that the Canonical coordinates have been taken to be 
connection components (and $\overline{W}_a$), resulting in the densitised 
components of the fundamental tensor appearing as conjugate momenta.
By discarding a total time derivative in the action, one could easily arrange 
for the fundamental tensor degrees of freedom to play their more traditional 
role as Canonical coordinates.}:
\begin{subequations}
\label{eq:momentum conditions}
\begin{align}
p&:=\frac{\delta{L_{\textsc{ngt}}}}{\delta[\mathrm{d}_{\mathsf{t}}u]}
=\sqrt{-\mathrm{g}},\\
p^a&:=\frac{\delta{L_{\textsc{ngt}}}}{\delta[\mathrm{d}_{\mathsf{t}}
\overline{W}_a]}=\mathbf{B}^a,\\
\pi^{ab}&:=\frac{\delta{L_{\textsc{ngt}}}}{\delta[\mathrm{d}_{\mathsf{t}}
{K}_{ab}]}=\pmb{\gamma}^{ab},
\end{align}
\end{subequations}
where the action has been written in terms of the Lagrangian as: 
$S_{\textsc{ngt}}=\int dt\,L_{\textsc{ngt}}$.
One could in principle treat all fields in the action as canonical coordinates, 
generating many cyclic momenta and hence many second class constraints that are 
trivially removed by the introduction of Dirac brackets \cite{Govaerts:1991}.
Instead the more economical approach of treating any field in the action whose 
time derivative does not appear anywhere as a Lagrange multiplier field will be 
followed\footnote{It is common to include the lapse and shift functions and 
their conjugate momenta in phase space, particularly if one is attempting to 
realize all of the generators of the spacetime diffeomorphism group on phase 
space \cite{Isham+Kuchar:1985a, Isham+Kuchar:1985b}.
This will not be done in this work simply to avoid the unnecessary algebra, as 
it is a trivial matter to extend the formalism later if necessary.}.

The extended Hamiltonian density is written as:
\begin{subequations}
\begin{equation}\label{eq:H}
{\mathcal H}_*={\mathcal H}_0
+N\phi_{ab}(\pi^{ab}-\pmb{\gamma}^{ab})
+2N\phi_a(p^a-\mathbf{B}^a)
+N\phi(p-\sqrt{-\mathrm{g}}),
\end{equation}
where the form of the Lagrange multiplier terms that enforce the algebraic 
conditions (\ref{eq:momentum conditions}) have been written as the reduction of 
the covariant form: $E\phi_{AB}(\pi^{AB}-\mathbf{g}^{AB})$.
The basic Hamiltonian density is derived from the Lagrangian density in the 
standard manner: $\mathcal{H}_0=\sum_IP^I\dot{Q}_I-\mathcal{L}_{\textsc{ngt}}$, 
and is found to be (equivalent to dropping time derivatives in (\ref{eq:L}) and 
taking its negative):
\begin{equation}
\begin{split}
{\mathcal H}_0=&
\mathcal{H}_m
+\sqrt{-\mathrm{g}}\pounds^{(3)}_{\vec{N}}[u] 
+\mathbf{B}^a\pounds^{(3)}_{\vec{N}}[\overline{W}]_a
+\pmb{\gamma}^{ab} \pounds^{(3)}_{\vec{N}}[K]_{ab}\\
&-N\sqrt{-\mathrm{g}}\bigl[
lv+l^a(b_a+\lambda^b_{ab})
+2L^a(c_a-a_a+{C_{\!\perp a}}^\perp)\\
&\quad\quad +\tfrac{1}{2}\alpha(W)^2
-v^a_{\;b}v^b_{\;a}
-\nabla^{(3)}_{a}[\sigma]^a
-\Gamma u
-\sigma^a(a_a-2c_a)
+u^a_{\;b}u^b_{\;a}\bigr]\\
&-N\mathbf{B}^a\bigl[
\nabla^{(3)}_{a}[W]+W(a_a-c_a)
-2\nabla^{(3)}_{b}[v]^b_{\;a}
+\nabla^{(3)}_{a}[v]  \\
&\quad\quad -2b_bu^b_{\;a}
+2b_au
-2(a_b-c_b)v^b_{\;a}+(a_a-c_a)v
-2j_{ab}\sigma^b
+2u^b_{\;c}\lambda^c_{ba}\bigr]\\
&-N\pmb{\gamma}^{[ab]}\bigl[
-\nabla^{(3)}_{{[a}}[\overline{W}]_{b]}
+\nabla^{(3)}_{c}[\lambda]^c_{ab}
+2\nabla^{(3)}_{{[a}}[b]_{b]} \\
&\quad\quad -j_{ac}u^c_{\;b}-j_{cb}u^c_{\;a}
+(\Gamma+u)j_{ab}
-k_{ca}v^c_{\;b}+k_{cb}v^c_{\;a}
+a_ab_b-a_bb_a+a_c\lambda^c_{ab}\bigr]\\
&-N\pmb{\gamma}^{(ab)}\bigl[
R^{(3)}_{(ab)}
-\nabla^{(3)}_{{(a}}[a]_{b)}
-a_aa_b  
+(\Gamma+u)k_{ab}
-k_{cb}u^c_{\;a}
-k_{ac}u^c_{\;b} \\
&\quad\quad-\tfrac{1}{2}\alpha(\overline{W}_a+\lambda^c_{\;ac}-b_a)
(\overline{W}_b+\lambda^d_{\;bd}-b_b)
+b_ab_b-j_{ac}v^c_{\;b}-j_{bc}v^c_{\;a}
+\lambda^d_{ac}\lambda^c_{bd}\bigr].
\end{split}
\end{equation}
\end{subequations}
Hamiltonians and Hamiltonian densities will be related everywhere by (for 
example): $H_*=\int_\Sigma d^3x\mathcal{H}_*$, where $H$ will refer to the 
Hamiltonian, and $\mathcal{H}$ to the related Hamiltonian density.

The canonical phase space variables that describe the configuration of the 
system at any time are:
\begin{equation}
\label{eq:variables}
\left\{(P^I,Q_I)\right\}=\left\{(p,u),(p^a,\overline{W}_a),
(\pi^{ab},K_{ab})\right\},
\end{equation}
and the usual canonical Poisson brackets have been assumed:
\begin{subequations}\label{eq:CCRs}
\begin{gather}
\{Q_I(x),Q_J(y)\}=\{P^I(x),P^J(y)\}=0,\\
\{Q_I[\mathbf{f}^I],P^J[g_J]\}=\int_\Sigma d^3z\,\mathbf{f}^I(z)g_I(z).
\end{gather}
\end{subequations}
The notation $Q_I[f^I]$ is used to denote the smearing of the object by an 
appropriately weighted test tensor (eg. $p^a[f_a]=\int_\Sigma d^3z 
p^a(z)f_a(z)$) leaving a scalar on $\Sigma$, and the sub/superscript `I' indexes 
the sets in (\ref{eq:variables}).  
The Poisson brackets are standard:
\begin{equation}\label{eq:Poisson}
\left\{F[Q,P],G[Q,P]\right\}=\int_\Sigma d^3z\sum_I\left[
\frac{\delta F}{\delta Q_I(z)}\frac{\delta G}{\delta P^I(z)}-\frac{\delta 
F}{\delta P^I(z)}\frac{\delta G}{\delta Q_I(z)}\right],
\end{equation}
where ($F,G$) are scalar functions of the canonical coordinates.

The fields $(p,u)$, although symmetric sector fields, do not show up in the 
standard Hamiltonian formulations of GR as canonical variables.
There one has $u=\gamma^{ab}k_{ab}$, and this is used in the action to either 
remove the time derivative from $u$ (second-order action) or write: 
$\partial_t{u}=\gamma^{ab}\partial_t{k}_{ab}+k_{ab}\partial_t{\gamma}^{ab}$ 
\cite{Gleiser+Holman+Neto:1987}.
The point being that in nonsymmetric theories, one cannot replace all terms in 
the action by various combinations of the spatial fundamental tensor and 
extrinsic curvature in a straightforward manner, and it seems one must be 
content to carry around the 103 Lagrange multipliers:
\begin{gather}
\{N\}:=\{N,N^a\},\quad
\{L\}:=\{(L^a),(l,l^a)\},\quad
\{\phi\}:=\{\phi,\phi_a,\phi_{(ab)},\phi_{[ab]}\},\nonumber\\
\{\gamma\}:=\{\mathbf{B}^a,\pmb{\gamma}^{ab}\},\quad
\{\Lambda\}:=\{W,b_a,v^a_{\;b},\lambda^a_{bc}\},\quad
\{\Gamma\}:=\{\Gamma,c_a,a_a,\sigma^a,w^a_{\;b},u^a_{Tb},\Gamma^a_{bc}\},
\end{gather}
where the Lagrange multipliers have been split up into smaller subsets for later 
reference.

\section{Constraint Analysis}
\label{sec:Constraint Analysis}

Now begins the analysis of the constraints and conditions that follow from 
variations of the Hamiltonian with respect to the Lagrange multipliers.
It will be found that all of the compatibility conditions and field equations 
(\ref{ncbtorsion},\ref{eq:compat},\ref{eq:mNGT field equations}) are reproduced 
from the Hamiltonian, leaving $\{N\}$ as undetermined Lagrange multipliers 
enforcing the diffeomorphism constraints in the standard form of the 
Hamiltonian.
Two second class constraints will also be found which could be used to strongly 
remove the conjugate pair $(p,u)$ from the phase space of the theory, leaving 
the symmetric phase space with the same dimension as in GR.
As the results are unsurprising and the algebra not unduly difficult but rather 
lengthy, the results will be given as sparsely as possible.

\subsection{Constraints Enforced by $\{\phi\}$, $\{L\}$, $\{\Lambda\}$ and 
$\{\Gamma\}$}
\label{sect:nonm}

The Lagrange multipliers $\{\phi\}$ are designed to enforce the conjugate 
momenta conditions (\ref{eq:momentum conditions}):
\begin{subequations}
\label{eq:momenta def}
\begin{align}
\label{eq:momenta def:gamma}
\frac{\delta{H}_*}{\delta\phi_{ab}}&\approx 
N\bigl[\pi^{ab}-\pmb{\gamma}^{ab}\bigr]\approx 0,\\
\label{eq:momenta def:B}
\frac{\delta{H}_*}{\delta\phi_a}&\approx 2N\bigl[p^a-\mathbf{B}^a\bigr]\approx 
0,\\
\label{eq:momenta def:density}
\frac{\delta{H}_*}{\delta\phi}&\approx 
N\bigl[p-\sqrt{-\mathrm{g}}\bigr]:=N\psi_1 \approx 0.
\end{align}
\end{subequations}
These conditions clearly determine $\{\gamma\}$, with one constraint left over: 
$\psi_1$, which has been identified in (\ref{eq:momenta def:density}). 
From $\{L\}$, the Lagrange multiplier enforcing the remaining torsion constraint 
from (\ref{Gamskew}) gives:
\begin{equation}
\label{eq:a relation}
\frac{\delta{H}_*}{\delta \mathbf{L}^a}\approx
-2N\bigl[c_a-a_a+{C_{\!\perp a}}^\perp\bigr]\approx 0,
\end{equation}
which will be used to determine $a_a$. 
The constraints resulting from the variation of $l^A$ are:
\begin{subequations}
\begin{align}\label{eq:la}
\frac{\delta{H}_*}{\delta \mathbf{l}}&\approx
-Nv\approx 0,\\
\frac{\delta{H}_*}{\delta \mathbf{l}^a}&\approx 
-N\bigl[b_a+\lambda^b_{ab}\bigr]\approx 0.
\end{align}
\end{subequations}
These are the components of (\ref{mNGTF:b}), and will be used to determine $v$ 
and $\lambda^b_{ab}$ respectively.

The conditions arising from the variation of $\{\Lambda\}$ are now calculated, 
first varying ${H}_*$ with respect to $W$ and then $v^b_{\;a}$:
\begin{subequations}
\begin{align}
\frac{\delta{H}_*}{\delta W}\approx &
N{\mathfrak C}^{[\perp a]}_ a\approx 0,\\
\frac{\delta{H}_*}{\delta v^b_{\;a}}\approx &
-2N\bigl[{\mathfrak C}^{[\perp a]}_b
-\tfrac{1}{3}\mathfrak{C}^{[\perp a]}_a\bigr]
-N\delta^a_b\bigl[\mathbf{l}-\tfrac{1}{3}\nabla^{(3)}_{c}[\mathbf{B}]^c\bigr]
\approx 0.
\end{align}
Using the result that $v\approx 0$ from (\ref{eq:la}), the trace of the last of 
these gives 
$\mathbf{l}\approx\frac{1}{3}\nabla^{(3)}_{c}[\mathbf{B}]^c
\approx\frac{\alpha}{3}\mathbf{W}$, so that $\mathfrak{C}^{[\perp a]}_b
\approx 0$.
Next varying $b_a$ gives:
\begin{equation}
\frac{\delta{H}_*}{\delta b_a}\approx
2N{\mathfrak C}^{[ab]}_b
-N\bigl[\mathbf{l}^a-\tfrac{1}{3}\alpha\pmb{\gamma}^{(ab)}W_b\bigr]\approx 0,
\end{equation}
and using the above conditions one finds that:
\begin{equation}
\frac{\delta{H}_*}{\delta \lambda^c_{ab}}\approx
-N{\mathfrak C}^{[ab]}_c\approx 0,
\end{equation}
\end{subequations}
which, combined with the previous result, yields 
$\mathbf{l}^a\approx\frac{\alpha}{3}\pmb{\gamma}^{(ab)}W_b$.
At this point, all of the skew sector algebraic compatibility conditions 
((\ref{eq:v relation}) and (\ref{eq:lambda relation})) as well as the field 
equations (\ref{mNGTF:d}) have been reproduced.

Varying $\{\Gamma\}$, one finds:
\begin{subequations}
\begin{align}
\label{eq:psi2}
\frac{\delta{H}_*}{\delta \Gamma}&\approx
N{\mathfrak C}^{(\perp a)}_a=:N\psi_2\approx 0,\\
\frac{\delta{H}_*}{\delta \sigma^a}&\approx
-N{\mathfrak C}^{(\perp \perp)}_ a\approx 0,\\
\frac{\delta{H}_*}{\delta c_a}&\approx
-2N\bigl[\mathbf{L}^a+\mathbf{\sigma}^a+\mathbf{B}^bv^a_{\;b}
-\tfrac{1}{2}\mathbf{B}^aW\bigr]\approx 0,
\end{align}
where the first of these yields a constraint (which has been identified as 
$\psi_2$ in (\ref{eq:psi2})) discussed further in Section \ref{sect:time evol}, 
and the last is used to determine the Lagrange multiplier $L^a$ as: 
\begin{equation}\label{eq:L defn}
L^a\approx-\sigma^a-B^bv^a_{\;b}+\frac{1}{2}B^aW.
\end{equation}
Using these, one computes:
\begin{equation}
\frac{\delta{H}_*}{\delta a_a}\approx
N\bigl[{\mathfrak C}^{(ab)}_b-{\mathfrak C}^{(\perp a)}_\perp\bigr]\approx 0,
\end{equation}
and the trace-free part of the variation with respect to $u^b_{\;a}$ gives the 
variation with respect to $u^b_{Ta}$:
\begin{equation} 
\frac{\delta{H}_*}{\delta u^b_{Ta}}\approx
-2N\bigl[{\mathfrak C}^{(\perp a)}_b-\tfrac{1}{3}\delta^a_b
{\mathfrak C}^{(\perp c)}_c\bigr]\approx 0,
\end{equation}
which yields $\mathfrak{C}^{(\perp a)}_b\approx 0$ when the constraint 
(\ref{eq:psi2}) is taken into account.
Finally, the variation with respect to the surface connection coefficients 
yields:
\begin{equation}
\frac{\delta{H}_*}{\delta \Gamma^c_{ab}}\approx
-N{\mathfrak C}^{(ab)}_c\approx 0,
\end{equation}
\end{subequations}
and all but $\mathfrak{C}_\Gamma$ of the symmetric sector algebraic 
compatibility conditions ((\ref{eq:sigma relation}), (\ref{eq:u relation}) and 
(\ref{eq:surface compatibility})) have been found, and the Lagrange multiplier 
$L^a$ determined by (\ref{eq:L defn}).

\subsection{Time Evolution of the Canonical Fields and the Constraints 
$\{\psi\}$}
\label{sect:time evol}

Time evolution of the canonical variables is determined as usual (from 
(\ref{eq:CCRs}) and (\ref{eq:Poisson})) by Hamilton's equations:
\begin{equation}\label{eq:can dot}
\partial_t{Q}_I=\{Q_I,{H}_*\}\approx \frac{\delta{H}_*}{\delta P^I},\quad
\partial_t{P}^I=\{P^I,{H}_*\}\approx -\frac{\delta{H}_*}{\delta Q_I}.
\end{equation}
These evolution equations should reproduce the same dynamics as given by  
(\ref{ncbtorsion}, \ref{eq:compat}, \ref{eq:mNGT field equations}).
It is convenient at this point to show that the canonical momenta evolve in 
accordance with the dynamical compatibility conditions, and that the evolution 
of the canonical coordinates also allows one to identify conditions that must be 
found in order to reproduce the NGT field equations (\ref{mNGTF:a}).
This is why this calculation is performed at this stage, even though strictly 
speaking it is not part of the constraint analysis.

The functional derivatives of ${H}_*$ with respect to the canonical coordinates 
will be necessary:
\begin{subequations}
\begin{align}
\frac{\delta{H}_*}{\delta u}&\approx
-\nabla^{(3)}_{a}[\mathbf{N}]^a
+N(\pmb{\Gamma}-\mathbf{u}-2\mathbf{B}^ab_a),\\
\frac{\delta{H}_*}{\delta\overline{W}_a}&\approx
-\mathbf{B}^a\nabla^{(3)}_{b}[N]^b
+\mathbf{B}^b\nabla^{(3)}_{b}[N]^a
-N^b\nabla^{(3)}_{b}[\mathbf{B}]^a\nonumber \\
&\quad\quad +N\bigl[\nabla^{(3)}_{b}[\pmb{\gamma}]^{[ab]}
+\pmb{\gamma}^{[ab]}{C_{\!\perp b}}^\perp
+\alpha \pmb{\gamma}^{(ab)}(\overline{W}_b-2b_b)\bigr],\\
\frac{\delta{H}_*}{\delta k_{ab}}&\approx
-\pmb{\gamma}^{(ab)}\nabla^{(3)}_{c}[N]^c
+\pmb{\gamma}^{(ac)}\nabla^{(3)}_{c}[N]^b
+\pmb{\gamma}^{(cb)}\nabla^{(3)}_{c}[N]^a
-N^c\nabla^{(3)}_{c}[\pmb{\gamma}]^{(ab)}\nonumber \\
&\quad\quad -N\bigl[\pmb{\gamma}^{(ab)}(\Gamma+u)
-2\mathbf{u}^{ab}\bigr],\\
\frac{\delta{H}_*}{\delta j_{ab}}&\approx
-\pmb{\gamma}^{[ab]}\nabla^{(3)}_{c}[N]^c
+\pmb{\gamma}^{[ac]}\nabla^{(3)}_{c}[N]^b
+\pmb{\gamma}^{[cb]}\nabla^{(3)}_{c}[N]^a
-N^c\nabla^{(3)}_{c}[\pmb{\gamma}]^{[ab]}\nonumber \\
&\quad\quad +N\bigl[\mathbf{B}^a\sigma^b-\mathbf{B}^b\sigma^a
-\pmb{\gamma}^{[ab]}(\Gamma+u)
+2\mathbf{v}^{ab}\bigr].
\end{align}
\end{subequations}
Time evolution of the conjugate momenta can be shown to be weakly equivalent to 
the dynamical compatibility conditions (\ref{eq:density dot}, 
\ref{eq:gamma dot}, \ref{eq:B dot}, \ref{eq:gamma skew dot}):
\begin{subequations}
\label{eq:metric evolution}
\begin{align}
\partial_t{p}&=-\frac{\delta{H}_*}{\delta u}
\approx\mathrm{d}_\mathsf{t}[\sqrt{-\mathrm{g}}]-N{\mathfrak 
C}^{\perp\perp}_\perp,\\
\partial_t{p}^a&=-\frac{\delta{H}_*}{\delta\overline{W}_a}
\approx\mathrm{d}_\mathsf{t}[\mathbf{B}]^a-N{\mathfrak C}^{[\perp a]}_\perp,\\
\partial_t{\pi}^{ab}&=-\frac{\delta{H}_*}{\delta K_{ab}}
\approx\mathrm{d}_\mathsf{t}[\pmb{\gamma}]^{ab}+N{\mathfrak C}^{ab}_\perp,
\end{align}
\end{subequations}
since in all cases (heuristically) 
$\partial_t{P}\approx\mathrm{d}_\mathsf{t}[\pmb{\gamma}]$, and one is left with 
the result that $\mathfrak{C}_\perp\approx 0$ in evolution.
Using (\ref{eq:field equations}) and the evolution of the coordinates:
\begin{subequations}
\label{eq:p dots}
\begin{align}
\partial_t{u}&=\frac{\delta{H}_*}{\delta p}
\approx N\phi,\\
\partial_t{\overline{W}}_a&=\frac{\delta{H}_*}{\delta p^a}
\approx 2N\phi_a,\\
\partial_t{K}_{ab}&=\frac{\delta{H}_*}{\delta \pi^{ab}}
\approx N\phi_{ab}, 
\end{align}
\end{subequations}
allows one to define the objects $\{\mathfrak{R}\}$:
\begin{subequations}
\label{eq:define phi}
\begin{align}
{\mathfrak R}&:= N\phi-\pounds^{(3)}_{\vec{N}}[u]-N{\mathfrak Z}_{\perp\perp}
+\tfrac{1}{4}m^2NM _{\perp\perp}, \\
{\mathfrak R}_a&:= 2N\phi_a-\pounds^{(3)}_{\vec{N}}[\overline{W}]_a
+2N{\mathfrak Z}_{[a\perp]}
-\tfrac{1}{2}m^2NM _{[a\perp]}, \\
{\mathfrak R}_{ab}&:= N\phi_{ab}-\pounds^{(3)}_{\vec{N}}[K]_{ab}
+N{\mathfrak Z}_{ab}
-\tfrac{1}{4}m^2NM _{ab}.
\end{align}
\end{subequations}
In order for Hamilton's equations (\ref{eq:can dot}) to reproduce the dynamical 
field equations (\ref{mNGTF:a}), one must find that $\{\mathfrak{R}\}\approx 0$, 
ensuring that $\{\phi\}$ are correctly determined. 

There are two constraints $\{\psi\}$ that have appeared from the variations so 
far ((\ref{eq:momenta def:density}) and (\ref{eq:psi2}) respectively):
\begin{subequations}\label{eq:psis}
\begin{align}
\label{eq:psis:1}
\psi_1&\approx p-\sqrt{-\mathrm{g}}[p^a,\pi^{ab}]\approx 0,\\
\label{eq:psis:2}
\psi_2&\approx pu-\pi^{ab}K_{ab}\approx 0,
\end{align}
\end{subequations}
where the density has been written as functionally dependent on the other 
momenta in order to stress that it must be considered as a functional of those 
fields alone (as in (\ref{eqn:variation density})) when calculating Poisson 
brackets.
Although all of the Lagrange multiplier variations have not been dealt with yet, 
the conditions that result from requiring that these constraints are preserved 
in time will prove useful at this stage.
Requiring that $\psi_1$ be preserved in time results in:
\begin{equation}
\partial_t{\psi}_1=
\{\psi_1,{H}_*\}\approx
-\frac{\delta{H}_*}{\delta u}
-\frac{2}{2-F}\beta_a\frac{\delta{H}_*}{\delta\overline{W}_a}
+\frac{1}{2-F}G_{ba}\frac{\delta{H}_*}{\delta K_{ab}}
\approx-\frac{4}{2-F}N{\mathfrak C}_\Gamma\approx 0,
\end{equation}
where (\ref{eqn:variation density}) has been used.
This directly gives the remaining algebraic compatibility condition 
(\ref{eq:Gamma relation}).
The preservation of $\psi_2$ gives:
\begin{equation}\label{eq:psi2 dot}
\partial_t{\psi}_2
=\{\psi_2,{H}_*\}\approx
\phi p-\pi^{ab}\phi_{ab}
-u\frac{\delta{H}_*}{\delta u}
+K_{ab}\frac{\delta{H}_*}{\delta K_{ab}}
\approx -\frac{4}{F-2}\sqrt{-\mathrm{g}}{\mathfrak R}
+2N{\mathfrak G}^\perp_{\;\perp},
\end{equation}
where $\mathfrak{R}$ is defined in (\ref{eq:define phi}) and 
$\mathfrak{G}^\perp_{\;\perp}$ by (\ref{eq:alg H}).   

\subsection{Variations of $\{\gamma\}$ and $\{N\}$}
\label{sect:phi}

In order to compute the variations with respect to $\{\gamma\}$, it is 
convenient to rewrite the extended Hamiltonian density (\ref{eq:H}) in the form 
($\{\mathfrak{Z}\}$ is defined in (\ref{eq:Z})):
\begin{multline}
{\mathcal H}_*\sim{\mathcal H}_m
+\sqrt{-\mathrm{g}}(N{\mathfrak 
Z}_{\perp\perp}+\pounds^{(3)}_{\vec{N}}[u]-N\phi)
+\mathbf{B}^a(-2N{\mathfrak 
Z}_{[a\perp]}+\pounds^{(3)}_{\vec{N}}[\overline{W}]_a-2N\phi_a)\\
+\pmb{\gamma}^{ab}(-N{\mathfrak 
Z}_{ab}+\pounds^{(3)}_{\vec{N}}[K]_{ab}-N\phi_{ab}),
\end{multline}
where $\sim$ refers to the fact that some conditions and constraints have been 
imposed in $\mathcal{H}_*$, provided that the results of the variations are not 
weakly affected.
(Note that this form holds for variations with respect to $\{\gamma\}$ 
{\em only}.)
The variation of $\mathcal{H}_m$ as defined in (\ref{eq:Hm}) can be written 
(once again making use of (\ref{eqn:variation density})):
\begin{equation}
\begin{split}
\delta{\mathcal H}_m&=-\tfrac{1}{4}Nm^2M _{AB}\delta\mathbf{g}^{AB}\\
&=-\frac{1}{2}Nm^2(M _{[\perp a]}
-\frac{1}{2-F}\beta_aM _{\perp\perp})\delta\mathbf{B}^a
+\frac{1}{4}Nm^2(M _{ab}
-\frac{1}{2-F}G_{ba}M _{\perp\perp})\delta\pmb{\gamma}^{ab},
\end{split}
\end{equation}
and one then computes the conditions enforced by $\{\gamma\}$ to be (using 
(\ref{eq:define phi})):
\begin{subequations}\label{eqn:metric vars}
\begin{align}
\frac{\delta{H}_*}{\delta \mathbf{B}^a}
&\approx-{\mathfrak R}_a+\frac{2}{2-F}\beta_a{\mathfrak R}\approx 0,\\
\frac{\delta{H}_*}{\delta \pmb{\gamma}^{ab}}
&\approx-{\mathfrak R}_{ab}-\frac{1}{2-F}\mathrm{G}_{ba}{\mathfrak R}\approx 0.
\end{align}
\end{subequations}

As it stands, the Hamiltonian $H_*$, is not in the standard form: 
$\int_\Sigma (N\mathcal{H}+N^a\mathcal{H}_a)$.
It may be put in this form by the removal of the surface term:
\begin{equation}\label{eq:Estar}
E_*:=\int_{\partial\Sigma}dS_a[
N^b\overline{W}_b \mathbf{B}^a
+2N^b\mathbf{k}^a_{\;b}
-2N\mathbf{L}^a],
\end{equation}
so that
\begin{equation}\label{Hngt}
{H}_*=\int_\Sigma d^3x(N{\mathcal H}+N^a{\mathcal H}_a)+E_*.
\end{equation}
(Further discussion of surface terms will occur in the next section.)
One then finds directly the Hamiltonian constraint ${\mathcal H}$:
\begin{subequations}\label{eq:diffeo constraints}
\begin{equation}
\label{eq:SH constraint}
{\mathcal H}:=\frac{\delta H_*}{\delta N}\approx -2{\mathfrak 
G}^\perp_{\;\perp}\approx 0,
\end{equation}
which has been weakly identified with the algebraic field equation 
(\ref{eq:alg H}) by making use of previous results.
This constraint, combined with (\ref{eq:psi2 dot}) and (\ref{eqn:metric vars}), 
establishes that $\{\mathfrak{R}\}\approx 0$. 
The momentum constraints are:
\begin{equation}\label{eq:SM constraint}
\begin{split}
{\mathcal H}_a&
:=\frac{\delta H_*}{\delta N^a}
= 2\mathbf{B}^b\nabla^{(3)}_{{[a}}[\overline{W}]_{b]}
-\overline{W}_a\nabla^{(3)}_{{b}}[\mathbf{B}]^b
+\pmb{\gamma}^{bc}\nabla^{(3)}_{{a}}[K]_{bc}
+\sqrt{-\mathrm{g}}\nabla^{(3)}_a[u]
-2\nabla^{(3)}_{{b}}[\mathbf{k}]^b_{\;a}\\
&\approx p^{b}e_{a}[\overline{W}_{b}]
-e_{b}[p^{b}\overline{W}_{a}]
+\pi^{bc}e_{a}[K_{bc}]
+pe_a[u]
-2e_{b}[\mathbf{k}^b_{\;a}]\\
&\approx -2{\mathfrak G}^\perp_{\;a}\approx 0,
\end{split}
\end{equation}
\end{subequations}
which has been identified with the related algebraic field equations 
(\ref{eq:alg Ha}).
The identity (\ref{eq:Mid}) has been used in order to make the identification of 
$\mathcal{H}$ and $\mathcal{H}_a$ with $\mathfrak{G}^\perp_{\;\perp}$ and 
$\mathfrak{G}^\perp_{\;a}$ respectively.
The form of the Hamiltonian constraint is not given explicitly in 
(\ref{eq:SH constraint}) as it can essentially be read off from (\ref{eq:H}) or 
(\ref{eq:alg H}) directly.
Unlike the case of the momentum constraint, there seems to be no relatively 
simple formulation of the Hamiltonian constraint in terms of canonical variables 
alone.
This leads to computational problems (as will be discussed in Section 
\ref{sect:CCA}), as Poisson brackets with $\mathcal{H}$ cannot be calculated in 
a straightforward manner.

With these last constraints, the Euler-Lagrange equations of NGT have been fully 
reproduced by Hamilton's equations combined with all of the algebraic 
compatibility conditions and field equations resulting from the variation of the 
Lagrange multipliers.
  
\subsection{Surface Terms}
\label{sect:surface terms}

Up to this point all surface terms have been consistently ignored.
This is fine if $\Sigma$ is closed ($\partial\Sigma=\emptyset$) and there is no 
surface to have terms on, but in general one must add a surface contribution to 
the action (\ref{eq:NGTAction}) designed to cancel off any variations of ${H}_*$ 
that do not vanish on the $\partial\Sigma$.
This ensures that the field equations that have been generated follow correctly 
from the Hamiltonian \cite{Regge+Teitelboim:1974}.

To begin with, one may add to $H_*$ the surface contribution:
\begin{equation}\label{eq:surface LM}
\begin{split}
E_{\text{\tiny{LM}}}=\int_{\partial\Sigma}dS_a[&
N\pi^{(bc)}\Gamma^a_{bc}
-N\pi^{(ab)}\Gamma^c_{bc}
-Np\sigma^a
-N\pi^{(ab)}a_b\\
&+N\pi^{[bc]}\lambda^a_{bc}
+2N\pi^{[ab]}b_b
-2Np^bv^a_{\;b}
+Np^av
+Np^aW],
\end{split}
\end{equation}
which ensures that the variation with respect to any of the determined Lagrange 
multipliers never result in any surface terms.
In addition, there are surface terms involving variations of $\{\gamma\}$ and 
$\{N\}$ to consider, resulting from variations of $E_{\text{\tiny{LM}}}$ as well 
as contributions from the variations computed earlier in this section.
These may be split up into terms involving the lapse and shift functions 
separately, and the full Hamiltonian for NGT including surface contributions 
written as:
\begin{equation}\label{eq:HNGT}
H_{\textsc{ngt}}:=\int_\Sigma d^3x(N\mathcal{H}+N^a\mathcal{H}_a)
+E_*+E_{\text{\tiny{LM}}}+E_{N}+E_{\vec{N}},
\end{equation}
where $E_*$ is given by (\ref{eq:Estar}).
The contribution from the shift function must satisfy:
\begin{subequations}
\label{eq:surface conditions} 
\begin{equation}\label{eq:surface shift}
\delta E_{\vec{N}}\approx -\int_{\partial\Sigma}dS_a[
N^ap\,\delta u
+N^a\pi^{bc}\delta K_{bc}
+\mathbf{k}^a_{\;b}\delta N^b
+N^ap^b\delta\overline{W}_b
+p^a\overline{W}_b\delta N^b],
\end{equation}
and is given solely in terms of Cauchy data and the undetermined Lagrange 
multipliers $\{N\}$.
The condition that the lapse contribution must satisfy is more complicated:
\begin{equation}\label{eq:surface lapse}
\begin{split}
\delta E_{N}\approx -\int_{\partial\Sigma}dS_a\big[&
N(\Gamma^a_{bc}\delta \pi^{(bc)}
-\Gamma^c_{bc}\delta\pi^{(ab)}
-\sigma^a\delta p
-a_b\delta\pi^{(ab)}
+\lambda^a_{bc}\delta\pi^{[bc]}
+\pi^{[ab]}\delta\overline{W}_b
+2b_b\delta\pi^{[ab]}
-2v^a_{\;b}\delta p^b
+W\delta p^a)\\
\quad &+(
\pi^{(bc)}\Gamma^a_{bc}
-\pi^{(ab)}\Gamma^c_{bc}
-p\sigma^a
-\pi^{(ab)}a_b
-2\mathbf{L}^a
+\pi^{[bc]}\lambda^a_{bc}
+2\pi^{[ab]}b_b
-2p^bv^a_{\;b}
+p^aW)\delta N\big].
\end{split}
\end{equation}
\end{subequations}
Note that this is a weak condition, and all determined Lagrange multipliers are 
properly expressed solely in terms of Cauchy data and $\{N\}$.
 
Just as is the case for GR, one cannot determine the form of these surface terms 
in general, but must be content to determine $E_{N}$ and $E_{\vec{N}}$ from 
(\ref{eq:surface conditions}) for each specific situation.
As a simple example, one may assume an asymptotically flat spacetime in which 
all of the antisymmetric functions fall off fast enough as $r\rightarrow\infty$ 
so that they will not contribute any surface terms.
Assuming the fall-off of the symmetric sector variables as given in 
\cite{Regge+Teitelboim:1974}, one finds that there are no additional 
contributions from (\ref{eq:surface conditions}).
Evaluating the non-vanishing surface terms: 
$\int_{\partial\Sigma}dS_a\,N\sqrt{-\mathrm{g}}(\gamma^{(bc)}\Gamma^a_{bc}
-\gamma^{(ab)}\Gamma^c_{bc})$, assuming the asymptotically Schwarzschild form 
given in \cite{DeWitt:1967}, yields the Schwarzschild mass parameter as 
expected.

\subsection{Completing the Constraint Analysis}
\label{sect:CCA}

It remains to be shown that the Hamiltonian and momentum constraints are 
preserved under time evolution (they lead to no further constraints), and that 
they can both be taken to be first class, leaving $\{\psi\}$ (\ref{eq:psis})  
second class.
It would then be possible to define Dirac brackets and impose $\{\psi\}$ 
strongly\footnote{This would be accomplished quite simply by replacing $u$ 
strongly everywhere using (\ref{eq:psis:2}), and considering both $p$ and 
$\sqrt{-\mathrm{g}}$ as functions of $\mathbf{B}^a$ and $\pmb{\gamma}^{ab}$, as 
determined by (\ref{eq:psis:1}) and (\ref{eqn:variation density}).
Thus $(p,u)$ would no longer be included in phase space, nor would they appear 
anywhere in the Hamiltonian.}. 
In order to examine the momentum constraints, it is useful to note (and can be 
explicitly checked) that they do indeed generate spatial diffeomorphisms on 
$\Sigma$:
\begin{equation}\label{eq:momentum generators}
\left\{Q_I,\mathcal{H}_a[f^a]\right\}\approx\pounds^{(3)}_{\vec{f}}[Q]_I,\quad
\left\{P^I,\mathcal{H}_a[f^a]\right\}\approx\pounds^{(3)}_{\vec{f}}[P]^I,
\end{equation}
and, as this may obviously be extended to any tensor field on $\Sigma$ built 
strictly out of the canonical variables as: 
$\left\{T,\mathcal{H}_c[f^c]\right\}\approx \pounds^{(3)}_{\vec{f}}[T]$, one 
finds (by explicit calculation, or using the above results for the last three):
\begin{subequations}
\begin{align}\label{eq:constraint algebra}
\left\{\psi_1[f_1],\psi_2[f_2]\right\}&\approx -\int_\Sigma 
d^3z\,f_1f_2p\frac{4}{2-F},\\
\left\{\psi_1[f],\mathcal{H}_a[f^a]\right\}&\approx -\int_\Sigma 
d^3z\,\psi_1\pounds^{(3)}_{\vec{f}}[f]\approx 0,\\
\left\{\psi_2[f],\mathcal{H}_a[f^a]\right\}&\approx -\int_\Sigma 
d^3z\,\psi_2\pounds^{(3)}_{\vec{f}}[f]\approx 0,\\
\left\{\mathcal{H}_a[f_1^a],\mathcal{H}_b[f_2^b]\right\}&\approx 
-\int_\Sigma d^3z\,\mathcal{H}_a\pounds^{(3)}_{\vec{f}_2}[f_1]^a\approx 0.
\end{align}
\end{subequations}
It has also been checked by explicit calculation that the time evolution of the 
momentum constraint (\ref{eq:SM constraint}) is weakly vanishing 
$\partial_t{\mathcal{H}}_a\approx 0$.

Thus the only condition required to show that the constraint algebra closes, is 
that $\left\{\mathcal{H}[M],\mathcal{H}[N]\right\}$ is some linear combination 
of the Hamiltonian and momentum constraints, and therefore vanishes weakly.
This is extremely difficult to check explicitly since (as noted before) 
$\mathcal{H}$ is not given in a simple form in terms of canonical variables 
alone, but instead depends on Lagrange multipliers which are difficult to solve 
for in general\footnote{Most of these Lagrange multipliers can in fact be 
ignored in Poisson brackets if $\mathcal{H}$ is kept in a form very similar to 
that appearing in $\mathcal{H}_*$, as their variations will vanish weakly using 
the algebraic compatibility conditions.
The variation of $\mathbf{L}^a$ (or $c_a$ if $a_a$ has been strongly removed 
from the system) will only vanish if the smearing function is exactly the lapse 
function $N$. 
This is not the case in general, and one would then require the variation of 
$b_a$ with respect to canonical variables in order to compute Poisson brackets 
containing $\mathcal{H}$; a variation that is not easy to compute.}.
This calculation however, need not be performed explicitly.

Considering the dynamics of NGT within the larger arena of hyperspace (the 
manifold of all spacelike embeddings \cite{Kuchar:1976a}), one may consider the 
surfaces $\Sigma_t$ as embedded in the Riemannian manifold 
$(\mathbf{M},\mathrm{g}^{(AB)})$.
A particular foliation defines a path in hyperspace, the tangent to which is the 
vector $\mathsf{t}$, as defined in Section \ref{Metric and Inverse}.
The hypersurfaces $\Sigma_t$ can then be viewed as deformations of an initial 
hypersurface $\Sigma_0$ along the vector field $\mathsf{t}$, and the Hamiltonian 
must then generate the evolution of the fields under these hypersurface 
deformations.
This implies that one {\em must} be able to write the Hamiltonian in the 
standard form: $\int_\Sigma d^3x\,(N\mathcal{H}+N^a\mathcal{H}_a)$ 
\cite{Isenberg+Nester:1980}. 
It can then be proved from the principle of path independence of dynamical 
evolution \cite{Teitelboim:1980}, that the Hamiltonian closing relations are 
constrained to take on the form:
\begin{equation}\label{eq:Ham closing}
\begin{split}
\left\{\mathcal{H}[M],\mathcal{H}[N]\right\}&=\int_\Sigma 
d^3z\,\mathcal{H}_a\gamma^{(ab)}(Me_b[N]-Ne_b[M]),\\
\left\{\mathcal{H}[M],\mathcal{H}_b[N^b]\right\}&= -\int_\Sigma 
d^3z\,\mathcal{H}\pounds^{(3)}_{\vec{N}}[M],\\
\left\{\mathcal{H}_a[M^a],\mathcal{H}_b[N^b]\right\}&= 
-\int_\Sigma d^3z\,\mathcal{H}_a\pounds^{(3)}_{\vec{N}}[M]^a.
\end{split}
\end{equation}
This principle essentially requires that the data on $\Sigma_0$ evolve uniquely 
to $\Sigma_t$, regardless of how $\Sigma_0$ is deformed into $\Sigma_t$ (i.e., 
independently of how one foliates the spacetime in-between), or equivalently, 
that evolution along different paths in hyperspace (between identical initial 
and final points) yield identical results.
This also ensures that the collection of all possible $\Sigma_t$'s is can be 
interpreted as describing different slicings of 
$(\mathbf{M},\mathrm{g}^{(AB)})$, and is therefore a reflection of the 
diffeomorphism invariance of spacetime\footnote{
In fact, a close inspection of the arguments in \cite{Teitelboim:1973} adapted 
to this case, shows that one only knows (\ref{eq:Ham closing}) up to a c-number 
(i.e. independent of the canonical variables).
This c-number has been assumed to be zero in this case, as it would not lead to 
further constraints, but to an inconsistent system of equations.
It is generally believed that the action for NGT generates a consistent set of 
field equations, and conservation laws (\ref{eq:NGT Bianchis}) have been derived 
based on diffeomorphism invariance that imply that this c-number vanishes.}.

The Hamiltonian for NGT is thus found to be of the form (\ref{eq:HNGT}), 
consistently generating  the field equations and compatibility conditions 
necessary to make Hamilton's equations (\ref{eq:can dot}) reproduce the field 
equations (\ref{ncbtorsion}, \ref{eq:compat}, \ref{eq:mNGT field equations}) 
derived from the action (\ref{eq:NGTAction}) via the Euler-Lagrange equations. 
The number of degrees of freedom for mNGT may now be easily computed using the 
standard algorithm as given by equation (1.60) of 
\cite{Henneaux+Teitelboim:1992}.
(For now it will be assumed that all of the Lagrange multipliers may be solved 
for, however this issue is discussed further in Section \ref{sect:Constraint 
problems}.)
In this case, one finds that the number of configuration space degrees of 
freedom per spacetime point is: the \# of canonical coordinates [$\{Q_I\}$ in 
(\ref{eq:variables})] ($13$) $-$ the \# of first class constraints 
[$\mathcal{H}$ and $\mathcal{H}_a$ in (\ref{eq:diffeo constraints})] ($4$) $-$ 
$\tfrac{1}{2}\times$ the \# of second class constraints [$\{\psi\}$ in 
(\ref{eq:psis})] ($\tfrac{1}{2}\times 2$) $=$ $8$.  
Since all of the constraints exist in the symmetric sector, two of these are the 
propagating modes of GR, and the remaining six occur in the NGT sector.
In the next section, the dynamics of pure GR configurations is reduced to that 
of GR (with a subtlety), and the limit to old NGT (Einstein's Unified Field 
Theory) is discussed.

\section{The reduction to GR and Old NGT}
\label{sec:limits}

It is worthwhile at this point to make further contact with GR by considering 
how one reduces NGT to GR configurations, as well as considering the dynamics of 
old NGT.

\subsection{GR}
\label{sec:GR}

There is, of course, an important surface in phase space on which all of the 
antisymmetric components of the fundamental tensor are identically zero.
If one sets up the system identically on this surface, it should remain on it 
since the dynamics there are identically those of GR (there is a complication, 
which will be discussed below).
One could easily just eliminate all skew objects from the action, and derive the 
ADM results directly.
In the context of NGT-type theories one would prefer to view the GR dynamics as 
occurring in the larger phase space of NGT, and view the reduction of dynamics 
as the imposition (by hand) of a constraint on the initial Cauchy surface that 
puts all skew Cauchy data weakly to zero.
Consistent dynamics on this surface will require that this constraint be 
preserved in time.

Considering first the case when $\alpha\neq 3/4$, one sets up fields on the 
initial surface such that all skew sector Cauchy data 
($p^a,\pi^{[ab]}, j_{ab}, \overline{W}_a$) vanish identically; the algebraic 
compatibility conditions require that $v^a_{\;b}=0$ and $\lambda^a_{bc}=b_a=0$ 
((\ref{eq:v relation}) and (\ref{eq:lambda relation}) respectively).
The time evolution conditions ((\ref{eq:B dot}) and (\ref{eq:gamma skew dot})) 
(or the equivalent in (\ref{eq:metric evolution})) guarantee that the 
antisymmetric components of the fundamental tensor ($B^a$ and $\gamma^{[ab]}$, 
respectively) will remain zero at later times.
The skew sector conjugate momenta vanish by the antisymmetric part of 
(\ref{eq:momenta def:gamma}) and (\ref{eq:momenta def:B}), and remain zero in 
evolution once  the previous results are inserted into (\ref{eq:p dots}) through 
the equations defining $\phi$ (\ref{eq:define phi}).  
The skew sector is thus consistently eliminated from any future dynamics once 
all skew Cauchy data is set to zero on the initial hypersurface.
The symmetric compatibility conditions (\ref{eq:symm compat}) reduce to those of 
GR, as do the symmetric time evolution equations in (\ref{eq:metric evolution})  
for (what is then unambiguously) the extrinsic curvature of the surface $k$.
The Hamiltonian and momentum constraints (\ref{eq:SH constraint}, \ref{eq:SM 
constraint}) then take on the familiar forms \cite{MTW:1973} 
($R^{(3)}=\gamma^{ab}R^{(3)}_{ab}$):
\begin{subequations}
\begin{align}\label{eq:GR SM}
\mathcal{H}&\approx 
\mathbf{k}^a_{\;b}k^b_{\;a}-\mathbf{k}^a_{\;a}k^b_{\;b}-\mathbf{R}^{(3)}
\approx -2\mathfrak{G}^\perp_{\perp}\approx 0,\\
\mathcal{H}_a&\approx -2\nabla_b[\mathbf{k}^b_{\;a}-\delta^b_a\mathbf{k}]
\approx -2\mathfrak{G}^\perp_{\;a}\approx 0.
\end{align} 
\end{subequations}

Thus given $\alpha\neq 3/4$, one finds that if at any time, all of the 
antisymmetric sector canonical variables vanish, then the ensuing dynamics of 
the system is identically that of GR.
The case where $\alpha= 3/4$ is slightly different, in that when one imposes 
(\ref{eq:lambda relation}) (and the remaining conditions), $b_a$ is left 
undetermined.
Then, although one still finds that $\partial_t{\pi}^{[ab]}\approx 0$ from 
(\ref{eq:metric evolution}), the evolution of $p^a$ is determined from 
$\partial_t{p}^a\approx\tfrac{3}{2}N\pi^{(ab)}b_b$, and is therefore 
undetermined.
Similar behavior occurs with the canonical coordinates $j_{ab},\overline{W}_a$, 
as there are nonvanishing contributions from $b_a$ to (\ref{eq:p dots}).
Thus the skew sector only remains trivial provided that $b$ is chosen to be zero 
on every surface, and the possibility exists that the system may pass through a 
GR configuration without remaining there, in contrast to the case 
above\footnote{Note that requiring 
$\partial_t{p}^a\approx\partial_t{\pi}^{[ab]}\approx 0$ on the initial surface 
is actually a stronger condition than $j_{ab},\overline{W}_a\approx 0$, as it 
implies that $b_a\approx 0$ initially.}.

\subsection{Old NGT}
\label{subsect:NGT}

The reduction to old NGT (or Einstein's Unified Field Theory) is accomplished by 
setting $\alpha=0$ (setting $m=0$ will not affect results, and can easily be 
left nonzero).
The algebraic compatibility conditions in this case may be solved for the 
determined Lagrange multipliers using similar methods to those in 
\cite{Tonnelat:1955,Tonnelat:1982,Tonnelat:1951}, with only mild conditions on 
the components of the fundamental tensor.
However (\ref{eq:W relation}) is now a constraint related to the $U(1)$ 
invariance that the action will now possess \cite{Moffat:1990}.
It is straightforward to show that this constraint is first class (its Poisson 
bracket with all other constraints is weakly vanishing) and generates $U(1)$ 
transformations on
$\Sigma$: $\{\overline{W}_a,e_b[p^b][\theta]\}\approx e_a[\theta]$.
There is then left five configuration space degrees of freedom per spacetime 
point, consistent with the results for the Unified Field Theory 
\cite{Lichnerowicz:1955} as well as NGT \cite{Moffat:1980, McDow+Moffat:1982} 
and \cite{Kunstatter+Leivo+Savaria:1984}.
In the last of these, {\it ad hoc} constraints are applied in Section 4 so as to 
remove $W_a$ from the dynamics, guaranteeing a weakly positive-definite 
Hamiltonian for the linearized theory.
These constraints cannot in fact be imposed even in the linearized theory 
\cite{Clayton:1995}, since lack of gauge invariance in the full action 
\cite{Kelly:1991,Kelly:1992} manifests itself as a non-conserved source for the 
skew tensor field, exciting all five modes even at the linearized level.
This is essentially equivalent to the results of Damour, Deser and McCarthy 
\cite{Damour+Deser+McCarthy:1993} who avoid discussing the source terms directly 
by demonstrating that the same effect is generated through the coupling of the 
skew field to a GR background. 

\section{A Closer Look at the Compatibility Conditions}
\label{sect:Constraint problems}

The symmetric sector algebraic constraints are easily solved for in terms of the 
skew sector Lagrange multipliers and Cauchy data.
$\Gamma$ is determined by (\ref{eq:Gamma relation}), 
$c_a$ is determined by (\ref{eq:c relation}), 
$a_a$ is determined by (\ref{eq:a relation}), 
$\sigma^a$ is determined by (\ref{eq:sigma relation}), 
$u^a_{\;b}$ is determined by (\ref{eq:u relation}), the trace of which is the 
constraint $\psi_2$. 
The relation (\ref{eq:surface compatibility}) can be used to uniquely solve for 
$\Gamma^a_{bc}$, provided the inverse of $\gamma^{(ab)}$ exists.
In the skew sector, $v^a_{\;b}$ is determined by (\ref{eq:v relation}), where 
the trace gives (\ref{eq:W relation}), determining the Lagrange multiplier $W$.
The remaining equations for $\lambda^a_{bc}$ and $b_a$: 
(\ref{eq:lambda relation}), are complicated since one must first replace the 
symmetric sector Lagrange multipliers, and then solve for the Lagrange 
multipliers in terms of Cauchy data.
The behavior relevant to this sector occurs in (\ref{eq:b relation}) alone, 
leaving the trace of (\ref{eq:lambda relation}) to determine $\lambda^a_{Tbc}$.
(\ref{eq:b relation}) can be written as an algebraic relation for $b_a$ solely 
in terms of Cauchy data (using (\ref{eq:c relation}) and (\ref{eq:u relation})):
\begin{subequations}
\begin{multline}\label{eq:full b}
p[(\tfrac{4}{3}\alpha-1)\gamma^{(ab)}+B^aB^b
-\tfrac{1}{3}\alpha\gamma^{[ac]}G_{[cd]}\gamma^{(db)}]b_b\\
\approx e_b[\pi^{[ab]}]
+k^b_{\;b}p^a-k^a_{\;b}p^b
+\tfrac{2}{3}\alpha\pi^{(ab)}\overline{W}_b
+\pi^{[ab]}j_{bc}B^c
-\tfrac{1}{6}\alpha\pi^{[ab]}\beta_bW
-\tfrac{1}{6}\alpha\pi^{[ab]}G_{[bc]}\gamma^{(cd)}\overline{W}_d,
\end{multline}
which will be written as (with obvious definitions):
\begin{equation}\label{eq:simple b}
pO^{ab}b_b\approx \mathbf{\Xi}^a.
\end{equation}
\end{subequations}

The issue at hand is whether or not the operator $O^{ab}$ is invertible in 
general.
Clearly if the skew sector is weak enough and $\alpha\neq 3/4$, then 
$\gamma^{(ab)}$ will dominate the operator, and $O^{-1}_{ab}$ will exist.
More generally there may exist only a subspace on which it is invertible, in 
which case the component(s) of $b_a$ in this space can be solved for, and the 
other(s) are left undetermined.
This will also imply that there are constraints corresponding to any 
undetermined components of $b_a$ of the form: $\mathbf{\Xi}^i\approx 0$.
One would then need to check that this constraint is preserved in time 
$\partial_t{\mathbf{\Xi}}^i\approx 0$, in order to determine whether it is 
possible for the system to remain on this constraint surface.
As discussed below in the case where the Cauchy data is dominated by the 
symmetric sector, it appears more likely that these are momentary configurations 
that the system may pass through.

For the case of mNGT ($\alpha=3/4$), one can easily see that the operator 
$O^{ab}$ is not dominated by the symmetric sector, and that in the absence of 
any antisymmetric components of the fundamental tensor, it disappears 
altogether.
Considering ($B^a,\gamma^{[ab]}$) as a perturbation on $\gamma^{(ab)}$, all 
quantities may be expanded to lowest order in powers of ($B^a, \gamma^{[ab]}$): 
\begin{gather}\label{eq:metric expansion}
\beta_a=B_a=\gamma_{(ab)}B^b,\quad
\alpha_a=-\gamma_{[ab]}B^b,\nonumber \\
\sqrt{-\mathrm{g}}=\sqrt{\gamma},\quad
G_{(ab)}=\gamma_{(ab)},\quad
G_{[ab]}=-\gamma_{[ab]}=-\gamma_{(ac)}\gamma_{(bd)}\gamma^{[cd]}.
\end{gather}
If one {\em assumes} that $b_a$ is of order ($B^a, \gamma^{[ab]}$) (indicated 
heuristically by $\mathcal{O}(\mathrm{skew})$), then (\ref{eq:simple b}) becomes 
three constraint equations $\mathbf{\Xi}^a\approx 0$.
These lead to $\partial_t{\mathbf{\Xi}}^a\approx 0$ for consistency, and one has 
obtained the dynamics of massive a Kalb-Ramond field on a GR background 
consistent with the linearized treatment of the field equations given in 
\cite{Moffat:1994,Moffat:1995b,Clayton:1995}.
This is not the general case, as one does not have the freedom to assume the 
behavior of the Lagrange multipliers, it must instead be derived from the 
compatibility conditions.
In particular, (\ref{eq:simple b}) determines $b_a$ in terms of Cauchy data, and 
keeping the dominant terms, one must find an inverse for the operator $O^{ab}$:
\begin{equation}\label{eq:O}
O^{ab}:= B^aB^b
+\tfrac{1}{4}\gamma^{[ac]}\gamma_{[cd]}\gamma^{(db)}
= B^aB^b-\gamma^{(ab)}\vec{\gamma}\cdot\vec{\gamma}+\gamma^a\gamma^b,
\end{equation}
where in the second form the following have been introduced:
\begin{equation}\label{eq:epsilon skew}
\gamma^{[ab]}=\frac{2}{\sqrt{\gamma}}\epsilon^{abc}\gamma_c,\quad 
\gamma_a:=\frac{1}{4}\sqrt{\gamma}\epsilon_{abc}\gamma^{[bc]},\quad
\gamma_{[ab]}=2\sqrt{\gamma}\epsilon_{abc}\gamma^c,\quad 
\gamma^a:=\frac{1}{4\sqrt{\gamma}}\epsilon^{abc}\gamma_{[bc]}.
\end{equation}
The Levi-Civita tensor density of weight $-1$: $\epsilon_{abc}$ and $+1$: 
$\epsilon^{abc}=\gamma^{(ad)} \gamma^{(be)} \gamma^{(cf)}\epsilon_{def}$ have 
been used ($\epsilon^{123}=\epsilon_{123}=+1$), and the notation 
$\vec{\gamma}\cdot\vec{\gamma}=\gamma^a\gamma_a$ has been introduced.
(Note that these are not quite duals; the extra numerical factor is introduced 
for convenience.)
The inverse of $O$ is found to be:
\begin{equation}\label{eq:inverse O}
O^{-1}_{ab}=-\frac{1}{\vec{\gamma}\cdot\vec{\gamma}}\left[\gamma_{(ab)}
-\frac{1}{\vec{\gamma}\cdot\vec{B}}(B_a\gamma_b+\gamma_aB_b)
+\frac{\vec{B}\cdot\vec{B}-\vec{\gamma}\cdot\vec{\gamma}}{(\vec{\gamma}\cdot\vec
{B})^2}\gamma_a\gamma_b\right].
\end{equation}
(Although the nonperturbative inversion of the operator $O^{ab}$ in 
(\ref{eq:simple b}) may be accomplished using similar techniques, the general 
result is not particularly enlightening and will not be given here.)
If $\vec{\gamma}\cdot\vec{\gamma}=0$ (in which case $\vec{\gamma}=0$if one 
assumes that $\gamma^{(ab)}$ is nondegenerate), then assuming that 
$\vec{B}\neq 0$, one finds two constraints coming from projecting perpendicular 
to $\vec{B}$, and one relation for $b_a$ from that parallel to $\vec{B}$. 
If $\vec{\gamma}\neq 0$ and either $\vec{B}=0$ or $\vec{\gamma}\cdot\vec{B}=0$ 
then there is one constraint from projecting parallel to $\vec{\gamma}$, and two 
conditions determining $b$ from the parallel component.
(The remaining operator when $\vec{B}=0$ projects out directions transverse to 
$\gamma$).
It has not been possible so far to find any of these cases which can be 
maintained in evolution, and so they are expected to be momentary configurations 
and not surfaces on which the system may evolve consistently.

The solution (\ref{eq:inverse O}) for $b_a$ implies that 
${\mathbf b}_a\approx O^{-1}_{ab}\mathbf{\Xi}^b$ is {\em not} 
$\mathcal{O}(\mathrm{skew})$, but in fact $\mathcal{O}(\mathrm{skew}^{-1})$.
(This does not happen in Einstein Unified Field Theory (or old NGT) since 
$\alpha=0$ leaves a term that allows one to solve for $b_a$ at lowest order.)
This in turn shows up in the evolution equations of the antisymmetric sector 
canonical variables ((\ref{eq:metric evolution}) and (\ref{eq:p dots})), causing 
them to evolve arbitrarily quickly as the skew initial data is made smaller.
It is possible to have $\Xi\approx\mathcal{O}(\mathrm{skew}^3)$, in which case 
$b$ will again reduce to $\mathcal{O}(\mathrm{skew})$, and this behavior is 
avoided.
However, this is a condition on the canonical variables alone, and since one 
could easily set up initial data that does not satisfy this condition, it would 
have to be realized dynamically.
At this point it is not known whether this is a reasonable expectation, though 
it seems unlikely that it would occur for {\em any} initial data configuration 
in which the skew sector is small\footnote{Note also that although one might 
expect that $\mathcal{H}$ might impose this constraint on the system due to the 
presence of terms of the form $\gamma^{(ab)}b_ab_a$ in $\mathcal{H}$, it can be 
shown that these terms identically cancel once $\alpha=3/4$ and 
$\lambda^a_{bc}=\lambda^a_{Tbc}+\delta^a_{[b}b_{c]}$ is employed.}.

This sort of behavior in the skew sector must be better understood, as the 
arbitrarily large time derivatives cause one to worry about whether GR 
spacetimes would be unstable in mNGT.
It is also not hard to see that the same situation may occur near NGT spacetimes 
as well.
As a specific example, consider the solution of (\ref{eq:full b}) specialized to 
spherically symmetric systems.
In the general case where both of ($B^1, \gamma^{[23]}$) are non-zero, the 
solution is:
\begin{equation}\label{eq:SS b}
b_1\approx 
\frac{1}{2}\frac{\gamma^{11}\overline{W}_1}{(B^1)^2}+2\frac{k^2_{\;2}}{B^1}.
\end{equation} 
Thus one finds that in regions of spacetime where $B^1$ is vanishingly small 
(for example,  perturbative situations, or in the asymptotic region of 
asymptotically flat spacetimes), $b_1$ becomes arbitrarily large, which in turn 
may cause ($\partial_t{p}^1, \partial_t{\overline{W}}_1, \partial_t{j}_{23}$) to 
become very large.
Although the Wyman sector solution \cite{Cornish:1994, Clayton:1995} (which is 
becoming the basis for most of the phenomenology in mNGT 
\cite{Moffat:1994, Legare+Moffat:1995b, Moffat+Sokolov:1995, Moffat:1995}) 
assumes that both $B^1$ and $\overline{W}_1$ vanish globally, if one considers 
perturbations of these fields on $\Sigma_0$, the above behavior reappears.

This is essentially the same effect as was found in \cite{Isenberg+Nester:1977}, 
where the effect of gravitational dynamics on the constraints of various 
derivative coupled vector fields was studied.
It was found that constraints on the vector field may be lost when GR is 
considered as evolving concurrently with the vector field (as opposed to the 
vector field evolving on a particular GR background).
This manifested itself as an increase in the number of degrees of freedom in the 
vector field, and singular behavior in the evolution equations when approaching 
asymptotically flat spacetimes.
There one finds no evidence of this when considering the vector field dynamics 
on a fixed GR background. 

This is intimately linked to degenerate solutions of the Euler-Lagrange 
equations.
If one considers linearized perturbations to such a solution, the field 
equations require that the perturbations vanish, whereas the analysis of the 
full dynamics near the degenerate solution may result in quite different 
behavior.
This degeneracy also results in the loss of hyperregularity (Chapter 7.4 of 
\cite{Marsden+Ratiu:1994}) for the Lagrangian system, and equivalence of the 
Hamiltonian system is not guaranteed.
However in this case, Hamilton's equations correctly reproduce the 
Euler-Lagrange equations provided the system avoids configurations corresponding 
to these degenerate solutions.

\section*{Conclusions}
\label{sect:disc}

The Hamiltonian formulation of both old (massless) NGT, as well as the massive 
theory have been given, identifying the former as having five antisymmetric 
sector configuration space degrees of freedom per spacetime point, and the 
latter six.
The symmetric sector of phase space has been enlarged over that of GR by a 
single canonical pair representing the fundamental tensor density 
$\sqrt{-\mathrm{g}}$. 
The existence of two related second class constraints allows one to remove these 
additional coordinates and recover the same number of degrees of freedom as GR 
in the symmetric sector.
The reduction of the system to pure GR field configurations has been discussed 
and the dynamics of GR recovered in the case of massless NGT.
In the massive theory, nontrivial dynamics of the antisymmetric sector occurs 
unless an undetermined Lagrange multiplier is chosen to vanish identically on 
every spacelike hypersurface. 

All fields have been decomposed by making use of a surface basis adapted to 
$\mathrm{g}^{(AB)}$, where the surface has been assumed to be spacelike (and the 
time vector timelike) with respect to all of the causal metrics of NGT 
$\{\mathrm{g}_c\}$.
Configuration space has been taken to consist of densitised components of the 
inverse of the fundamental tensor for convenience, broken up into the lapse and 
shift functions and the surface degrees of freedom: ($N, N^a, 
\pmb{\gamma}^{(ab)},\mathbf{B}^a, \pmb{\gamma}^{[ab]}$).
The lapse and shift functions remain undetermined Lagrange multipliers, and 
phase space consists of canonical pairs that include all of the remaining 
components of the inverse of the fundamental tensor, as well as an additional 
redundant pair mentioned above.
The Hamiltonian has been put into standard form, and the lapse and shift 
functions enforce the NGT Hamiltonian and momentum constraints ($\mathcal{H}$ 
and $\mathcal{H}_a$ respectively).

The Hamiltonian for massive NGT has been found to contain determined Lagrange 
multipliers that become singular near particular field configurations.
This may occur when the system is very close to any GR field configuration, and 
has been noted to also occur for Cauchy data near the asymptotically 
well-behaved static spherically symmetric solution with nontrivial NGT sector 
(the massive Wyman solution).
This behavior is not uncommon in derivative coupled theories, where the rank of 
the kinetic matrix changes for particular field configurations.
At this point in time the consequences of this behavior for the massive theory 
are not well understood.

\section*{Acknowledgements}

The author would like to thank the Natural Sciences and Engineering Research 
Council of Canada and the University of Toronto for funding during part of this 
work.
Thanks also to J. W. Moffat, P. Savaria, L. Demopoulos, and N. Cornish for 
discussions related to the work, and especially to J. L\'{e}gar\'{e} for a 
critical reading of the manuscript.

\appendix

\section{Surface Decomposed Compatibility Conditions and Field Equations}

The surface decomposition of (\ref{eq:compat}) and (\ref{mNGTF:a}) is 
straightforward, however as these will be required weakly, the algebraic 
conditions that determine the Lagrange multipliers, as well as the remaining 
torsion-free condition, have been applied to write them in the given form.
The definition of $\overline{W}_a$ given by (\ref{eq:Wbar}) has also been used 
throughout. 

\subsection{Compatibility Conditions}
\label{sec:compat}

The symmetric components of (\ref{eq:compat}) are found to be:
\begin{subequations}\label{eq:symm compat}
\begin{align}
\label{eq:density dot}
{\mathfrak C}_\perp^{(\perp\perp)}&=
\mathrm{d}_{e_\perp}[\sqrt{-\mathrm{g}}]
+\pmb{\Gamma}-\mathbf{u}
-2\mathbf{B}^ab_a,  \\
\label{eq:gamma dot}
{\mathfrak C}_\perp^{(ab)}&=
-\mathrm{d}_{e_\perp}[\pmb{\gamma}]^{(ab)}
+\pmb{\gamma}^{(ab)}(\Gamma+u)
-2\mathbf{u}^{ab},  \\
\label{eq:sigma relation}
{\mathfrak C}_\perp^{(\perp a)}&=
\mathbf{\sigma}^a
-\pmb{\gamma}^{(ab)}a_b
+\mathbf{B}^bv^a_{\;b}+\pmb{\gamma}^{[ab]}b_b,  \\
{\mathfrak C}_a^{(\perp\perp)}&=
\nabla^{(3)}_{a}[\sqrt{-\mathrm{g}}]
+\mathbf{c}_a
+2 j_{ab}\mathbf{B}^b,  \\
\label{eq:u relation}
{\mathfrak C}_b^{(\perp a)}&=
\mathbf{u}^a_{\;b}
-\mathbf{B}^ab_b
-\mathbf{B}^c\lambda^a_{cb}
-\mathbf{k}^a_{\;b},  \\
\label{eq:surface compatibility}
{\mathfrak C}_c^{(ab)}&=
-\nabla^{(3)}_{c}[\pmb{\gamma}]^{(ab)}
+\pmb{\gamma}^{(ab)}c_c
+\mathbf{B}^av^b_{\;c}
+\mathbf{B}^bv^a_{\;c}
+\pmb{\gamma}^{[ad]}\lambda^b_{dc}
+\pmb{\gamma}^{[db]}\lambda^a_{cd},
\end{align}
\end{subequations}
and the skew components:
\begin{subequations}
\begin{align}
\label{eq:B dot}
{\mathfrak C}_\perp^{[\perp a]}&=
\mathrm{d}_{e_\perp}[\mathbf{B}]^a
-\mathbf{B}^au+\mathbf{B}^bu^a_{\;b}
+\pmb{\gamma}^{[ab]}a_b
-(1+\tfrac{2}{3}\alpha)\pmb{\gamma}^{(ab)}b_b
+\tfrac{1}{3}\alpha\pmb{\gamma}^{(ab)}\overline{W}_b,  \\
\label{eq:gamma skew dot}
{\mathfrak C}_\perp^{[ab]}&=
-\mathrm{d}_{e_\perp}[\pmb{\gamma}]^{[ab]}
-\mathbf{B}^a\sigma^b+\mathbf{B}^b\sigma^a
+\pmb{\gamma}^{[ab]}(\Gamma+u)
-2\mathbf{v}^{ab},  \\
\label{eq:v relation}
{\mathfrak C}_b^{[\perp a]}&=
\nabla^{(3)}_{b}[\mathbf{B}]^a
-\mathbf{v}^a_{\;b}
+\mathbf{j}^a_{\;b}
-\tfrac{1}{3}\alpha\delta^a_b\mathbf{W},  \\
\label{eq:lambda relation}
{\mathfrak C}_c^{[ab]}&=
-\nabla^{(3)}_{c}[\pmb{\gamma}]^{[ab]}
-\mathbf{B}^au^b_{\;c}
+\mathbf{B}^bu^a_{\;c}
+\pmb{\gamma}^{[ab]}c_c
+\pmb{\gamma}^{(ad)}\lambda^b_{dc}
+\pmb{\gamma}^{(db)}\lambda^a_{cd}
+\tfrac{2}{3}\alpha\delta^{[a}_c\delta^{b]}_d\pmb{\gamma}^{(de)}
(\overline{W}_e-2b_e).  
\end{align}
\end{subequations}
The first two of each of these are dynamical compatibility conditions, and the 
rest are algebraic conditions that determine Lagrange multipliers.
It is useful to have the contractions of a few of these:
\begin{subequations}
\begin{align}
{\mathfrak C}^{(\perp a)}_a&=\mathbf{u}-\mathbf{k}^a_{\;a},\\
{\mathfrak C}^{(ab)}_b&=-\nabla^{(3)}_{b}[\pmb{\gamma}]^{(ab)}
+\pmb{\gamma}^{(ab)}c_b
+\mathbf{B}^bv^a_{\;b}
-\pmb{\gamma}^{[ab]}b_b
+\pmb{\gamma}^{[bc]}\lambda^a_{cb},\\
\label{eq:W relation}
{\mathfrak C}^{[\perp a]}_a&=
\nabla^{(3)}_{a}[\mathbf{B}]^a-\alpha\mathbf{W},\\
\label{eq:b relation}
{\mathfrak C}^{[ab]}_b&=-\nabla^{(3)}_{b}[\pmb{\gamma}]^{[ab]}
-\mathbf{B}^au+\mathbf{B}^bu^a_{\;b}
+\pmb{\gamma}^{[ab]}c_b
+(\tfrac{4}{3}\alpha-1)\pmb{\gamma}^{(ab)}b_b
-\tfrac{2}{3}\alpha\pmb{\gamma}^{(ab)}\overline{W}_b.
\end{align}
\end{subequations}

Note that there is one further algebraic compatibility condition buried in the 
dynamical compatibility conditions.
By taking appropriate combinations in order to make use of (\ref{eqn:variation 
density}) (replacing the variation with a derivative off the surface), one can 
find:
\begin{subequations}
\begin{equation}
\label{eq:Gamma relation}
{\mathfrak C}_\Gamma:=\pmb{\Gamma}
-\mathbf{B}^ab_a
+\tfrac{1}{6}\alpha\pmb{\gamma}^{(ab)}\beta_a(\overline{W}_b-2b_b),
\end{equation}
giving the remaining algebraic condition that determines $\Gamma$.
Repeating this last calculation and considering a derivative in the surface 
yields a condition that determines $c_a$:
\begin{equation}
\label{eq:c relation}
{\mathfrak C}_c:=\mathbf{c}_a+j_{ab}\mathbf{B}^b
-\tfrac{1}{6}\alpha\beta_a\mathbf{W}
-\tfrac{1}{6}\alpha G_{[ab]}\pmb{\gamma}^{(bc)}(\overline{W}_c-2b_c),
\end{equation}
\end{subequations}
although this is not independent of the algebraic conditions already found.

\subsection{Algebraic Field Equations and Conservation Laws}
\label{sec:feq}

The decomposition of the field equations (\ref{mNGTF:a}) yields:
\begin{subequations}
\label{eq:field equations}
\begin{align}
{\mathcal R}_{\perp\perp}&=-\mathrm{d}_{e_\perp}[u]+{\mathfrak Z}_{\perp\perp}
-\tfrac{1}{4}m^2M _{\perp\perp},\\
\label{eq:Raperp}
{\mathcal R}_{(a\perp)}&=-\mathrm{d}_{e_\perp}[j_{ab}B^b]
+{\mathfrak Z}_{(a\perp)}-\tfrac{1}{4}m^2M _{(a\perp)},\\
{\mathcal 
R}_{[a\perp]}&=\tfrac{1}{2}\mathrm{d}_{e_\perp}[\overline{W}]_a
+{\mathfrak Z}_{[a\perp]}
-\tfrac{1}{4}m^2M _{[a\perp]},\\
{\mathcal R}_{ab}&=\mathrm{d}_{e_\perp}[K]_{ab}+{\mathfrak Z}_{ab}
-\tfrac{1}{4}m^2M _{ab},
\end{align}
\end{subequations}
the second of which has made use of the Jacobi identity (\ref{eq:Jacobi}).
For completeness, the components of the mass tensor are:
\begin{subequations}
\begin{align}
M _{\perp\perp}&=
F(F-1)-2\gamma^{(ab)}\beta_a\beta_b+\tfrac{1}{2}F\gamma^{[ab]}G_{[ab]},\\
M _{(a\perp)}&=
-\tfrac{1}{2}\gamma^{[bc]}G_{[bc]}\alpha_a+\beta_c(G_{[ab]}\gamma^{(bc)}
-G_{(ab)}\gamma^{[bc]}),\\
M _{(ab)}&=
-(F-1)G_{(ab)}-\tfrac{1}{2}G_{(ab)}\gamma^{[cd]}G_{[cd]}+2\beta_a\beta_b
-2\alpha_a\alpha_b
+\tfrac{1}{2}\gamma^{[cd]}(G_{ca}G_{bd}-G_{db}G_{ac}),\\
M _{[a\perp]}&=
-\beta_a+\tfrac{1}{2}\gamma^{[bc]}G_{[bc]}\beta_a
-\beta_c(G_{(ab)}\gamma^{(bc)}-G_{[ab]}\gamma^{[bc]}),\\
M _{[ab]}&=
(F-2)G_{[ab]}+\tfrac{1}{2}G_{[ab]}\gamma^{[cd]}G_{[cd]}+2\beta_a\alpha_b
-2\beta_b\alpha_a
-\tfrac{1}{2}\gamma^{[cd]}(G_{cb}G_{ad}-G_{ca}G_{bd}),
\end{align}
\end{subequations}
for which a few identities exist:
\begin{subequations}\label{eq:Mid}
\begin{align}
M _{\perp\perp}+\gamma^{ab}M _{ab}&=-2(F-1)-\gamma^{[cd]}G_{[cd]},\\
M _{(a\perp)}+ M _{[ab]}B^b&=0,\\
\gamma^{(ab)}M _{(\perp b)}+\gamma^{[ab]}M _{[\perp b]}&=0.
\end{align}
\end{subequations}
The remaining objects in (\ref{eq:field equations}) $\{\mathfrak{Z}\}$ are:
\begin{subequations}\label{eq:Z}
\begin{align}
{\mathfrak Z}_{\perp\perp}&:=
\nabla^{(3)}_{a}[\sigma]^a+\Gamma u-u^a_{\;b}u^b_{\;a}+\sigma^a(a_a-2c_a)
+v^a_{\;b}v^b_{\;a}-\tfrac{1}{2}\alpha(W)^2,\\
{\mathfrak Z}_{(a\perp)}&:=
-\nabla^{(3)}_{a}[u]
+\nabla^{(3)}_{b}[u]^b_{\;a}
-\nabla^{(3)}_{a}[B^bb_b]\nonumber \\
&\quad\quad +uc_a-k_{ab}\sigma^b
-(a_a-c_a)B^bb_b+u^b_{\;a}(a_b-c_b)
+v^b_{\;a}b_b+v^b_{\;c}\lambda^c_{ab}+\tfrac{1}{2}\alpha 
W(\overline{W}_a-2b_a),\\
{\mathfrak Z}_{(ab)}&:=
R^{(3)}_{(ab)}-\nabla^{(3)}_{{(a}}[a]_{b)}+(\Gamma+u)k_{ab}
-a_aa_b-k_{cb}u^c_{\;a}-k_{ac}u^c_{\;b} \nonumber \\
&\quad\quad +b_ab_b
-j_{ac}v^c_{\;b}-j_{bc}v^c_{\;a}+\lambda^c_{ad}\lambda^d_{bc}
-\tfrac{1}{2}\alpha(\overline{W}_a-2b_a)(\overline{W}_b-2b_b),\\
{\mathfrak Z}_{[a\perp]}&:=
\frac{1}{2}\nabla^{(3)}_{a}[W]+\frac{1}{2}W(a_a-c_a)
-\nabla^{(3)}_{b}[v]^b_{\;a}-j_{ab}\sigma^b +(c_b-a_b)v^b_{\;a}
+ub_a-u^b_{\;a}b_b-u^b_{\;c}\lambda^c_{ab},\\
{\mathfrak Z}_{[ab]}&:=
-\nabla^{(3)}_{[a}[\overline{W}]_{b]}+2\nabla^{(3)}_{[a}[b]_{b]}
+\nabla^{(3)}_{c}[\lambda]^c_{ab}\nonumber \\
&\quad\quad +(\Gamma+u)j_{ab}
+a_c\lambda^c_{ab}+a_ab_b-a_bb_a-j_{cb}u^c_{\;a}-j_{ac}u^c_{\;b}
-k_{ca}v^c_{\;b}+k_{cb}v^c_{\;a}.
\end{align}
\end{subequations}

One may construct the tensor density (these are, of course, just particular 
combinations of the field equations):
\begin{equation}
\mathfrak{G}^A_{\;B}:=\tfrac{1}{2}({\mathbf g}^{AC}{\mathcal G}_{BC}
+{\mathbf g}^{CA}{\mathcal G}_{CB})
=\tfrac{1}{2}({\mathbf g}^{AC}{\mathcal R}_{BC}
+{\mathbf g}^{CA}{\mathcal R}_{CB}
-\delta^A_B{\mathbf g}^{CD}{\mathcal R}_{CD}),
\end{equation}
where ${\mathcal G}_{AB}={\mathcal 
R}_{AB}-\tfrac{1}{2}\mathrm{g}_{BA}\mathcal{R}$ is the equivalent of the 
Einstein tensor for NGT.
It is a fairly straightforward calculation to show that the components 
$\mathcal{G}^\perp_{\;\perp}$ and $\mathcal{G}^\perp_{\;a}$ are both algebraic 
field equations:
\begin{subequations}
\begin{align}
\label{eq:alg H}
{\mathfrak G}^\perp_{\;\perp}
&=\tfrac{1}{2}\bigl[\sqrt{-\mathrm{g}}{\mathcal 
R}_{\perp\perp}+\pmb{\gamma}^{ab}{\mathcal R}_{ab}\bigr]\nonumber \\
&=\tfrac{1}{2}\bigl[\sqrt{-\mathrm{g}}{\mathfrak 
Z}_{\perp\perp}+\pmb{\gamma}^{ab}{\mathfrak Z}_{ab}
-\tfrac{1}{4}m^2(\sqrt{-\mathrm{g}}M _{\perp\perp}+\pmb{\gamma}^{ab}M 
_{ab})\bigr]
-\Gamma\mathbf{u}+u\mathbf{B}^ab_a+j_{ab}\mathbf{B}^a\sigma^b
+k^a_{\;b}\mathbf{u}^b_{\;a}-j^a_{\;b}\mathbf{v}^b_{\;a},\\
\label{eq:alg Ha}
\mathfrak{G}^\perp_{\;a}&=\sqrt{-\mathrm{g}}{\mathcal 
R}_{(a\perp)}+\mathbf{B}^b{\mathcal R}_{[ab]}\nonumber \\
&=\sqrt{-\mathrm{g}}{\mathfrak Z}_{(a\perp)}+{\mathfrak Z}_{[ab]}\mathbf{B}^b
+j_{ab}\bigl[
-\pmb{\Gamma}B^b+u^b_{\;c}\mathbf{B}^c
+2B^b\mathbf{B}^cb_c
+\pmb{\gamma}^{[bc]}a_c
-(1+\tfrac{4}{3}\alpha)\pmb{\gamma}^{(bc)}b_c
+\tfrac{2}{3}\alpha\pmb{\gamma}^{(bc)}\overline{W}_c\bigr],
\end{align}
\end{subequations}
analogous to the Gauss and Codacci relations respectively in GR.
The conservation laws for NGT derived in \cite{Legare+Moffat:1995} written in a 
general basis are:
\begin{equation}\label{eq:NGT Bianchis}
\mathcal{B}_A:=\nabla_{e_B}[\mathfrak{G}]^B_{\;A}
+\tfrac{1}{2}\mathcal{R}_{BC}\nabla_{e_A}[\mathbf{g}]^{BC}=0,
\end{equation}
guaranteeing the existence of four Bianchi-type identities related to the above 
algebraic constraints.
Note that these identities cannot be used here in this form, as the 
compatibility conditions have been imposed in their derivation.
Although this will not be necessary here, one could in principle derive the form 
of (\ref{eq:NGT Bianchis}) that would hold without imposing the compatibility 
conditions, and could therefore be used here to derive the closing relations 
(\ref{eq:Ham closing}) directly from the action (\ref{eq:NGTAction}); using the 
assumption of diffeomorphism invariance in a slightly more operational manner. 

\newpage

\bibliographystyle{amsplain}

\providecommand{\bysame}{\leavevmode\hbox to3em{\hrulefill}\thinspace}

\end{document}